\documentclass[fleqn,usenatbib]{mnras}

\usepackage{newtxtext,newtxmath}

\usepackage[T1]{fontenc}

\DeclareRobustCommand{\VAN}[3]{#2}
\let\VANthebibliography\thebibliography
\def\thebibliography{\DeclareRobustCommand{\VAN}[3]{##3}\VANthebibliography}

\usepackage{multirow}
\usepackage{graphicx}

\usepackage{graphicx}	
\usepackage{amsmath}	

\title[Time-Scales]{Exploring Galaxy Evolution Time-Scales in Clusters: Insights from the Projected Phase Space}

\author[Sampaio et al. 2023]{V. M. Sampaio,$^{1,2}$\thanks{E-mail: vitorms999@gmail.com}
R. R. de Carvalho,$^{1}$ A. Arag\'on-Salamanca,$^{2}$ M. R. Merrifield$^{2}$, I. Ferreras,$^{3,4}$ \newauthor D. J. Cornwell$^{2}$   
\\
$^{1}$ NAT - Universidade Cidade de S\~ao Paulo, 01506-000, SP, Brazil\\
$^{2}$ School of Physics and Astronomy, University of Nottingham, University Park, Nottingham NG7 2RD, UK\\
$^{3}$ Instituto de Astrof\'isica de Canarias, Calle V\'ia L\'actea s/n,
E38205, La Laguna, Tenerife, Spain\\
$^{4}$ Department of Physics and Astronomy, University College London, Gower Street, London WC1E 6BT, UK
}

\date{Accepted 2024 June 17. Received 2024 June 17; in original form 2024 February 29}

\pubyear{2024}

\begin{document}
\label{firstpage}
\pagerange{\pageref{firstpage}--\pageref{lastpage}}
\maketitle

\begin{abstract}
Galaxies infalling into clusters undergo both star-formation quenching and morphological transformation due to environmental effects. We investigate these processes and their timescales using a local sample of 20,191 cluster and 11,674 field galaxies from SDSS. By analysing morphology as a function of distance from the star-formation main sequence, we show that environmental influence is especially pronounced for low-mass galaxies, which emerge from the green valley with early-type morphologies before their star formation is fully suppressed. Using the galaxies' positions in the clusters' Projected Phase Space, we examine the evolution of blue cloud, green valley, and red sequence fractions as a function of time since infall. Interestingly, the green valley fraction remains constant with time since infall, suggesting a balanced flow of galaxies in and out of this class. We estimate that galaxies less massive than $10^{10}\rm M_{\odot}$ spend approximately 0.4 Gyr in the green valley. By comparing quenched and early-type populations, we provide further evidence for the ``slow-then-rapid'' quenching model and suggest that it can also be applied to morphological transitions. Our results indicate that morphological transformation occurs at larger radii than complete star-formation quenching. About 75\% of galaxies undergoing morphological transition in clusters are spirals evolving into S0s, suggesting that infalling galaxies retain their disks, while massive ellipticals are relics of early merger events. Finally, we show it takes approximately 2.5 and 1.2 Gyr after the delay-time ($\sim 3.8 {\rm Gyr}$) for the population of low mass galaxies in clusters to reach a 50\% threshold in quenched and early-type fraction, respectively. These findings suggest morphological transition precedes full star formation quenching, with both processes possibly being causally linked.
\end{abstract}




\section{Introduction}

Galactic systems exhibit a diverse range of properties, influenced by gravitational and hydrodynamic interactions with their surrounding environments. The study by \citet{Dressler} offers initial insights into the influence of environment on galactic morphology through the Morphology-Density relation. In brief, early-type galaxies prevail in high-density environments, whereas spiral galaxies are predominantly found in low-density fields. This pattern is explained by enhanced gravitational interactions, intra-cluster medium pressure, and encounters with neighboring galaxies \citep{2019igfe.book.....C}.

The interplay between mechanisms affecting both star formation and morphology in galaxies prompts the investigation of the balance between ``internal'' and ``environmental'' effects. The former are predominantly related to intrinsic galaxy properties. Notably, Active Galactic Nuclei (AGN) feedback \citep{2008MNRAS.387.1431D, Bongiorno, 2020MNRAS.491.5406T, 2023MNRAS.524.5327S}, ``Morphological Quenching'' \citep{2009ApJ...707..250M}, and bar-driven evolution \citep{2017MNRAS.465.3729S} are just a few internal mechanisms. Conversely, environmental quenching encompasses processes dependent on environmental density and halo mass, such as ``starvation'' \citep{1980ApJ...237..692L,2000ApJ...540..113B,2017MNRAS.466.3460V, 2020MNRAS.491.5406T}, ``tidal-mass-loss'' \citep{1999MNRAS.302..771J,2006MNRAS.366..429R}, and ram-pressure stripping (RPS) \citep{1972ApJ...176....1G,1999MNRAS.308..947A}. Additionally, major and minor mergers significantly impact the morphology and composition of satellite galaxies \citep{2005ApJ...622L...9S,2008MNRAS.384..386C,2010ApJ...720L.149T}. Large-scale structures, primarily through filaments, can also induce notable galactic evolution prior to cluster infall, a phenomenon termed ``pre-processing'' \citep{1998ApJ...496...39Z, 2004PASJ...56...29F,2013MNRAS.431L.117M,2019A&A...632A..49S, 2021MNRAS.503.3065S}.

A significant finding by \citet{2010ApJ...721..193P} demonstrates that these internal and external drivers can be disentangled for galaxies up to redshift $\sim 1$. This results from the high density in clusters being more effective at gas removal and star formation suppression in low-mass galaxies \citep{2010ApJ...721..193P, 2017ApJ...843..128R, 2018MNRAS.476.1680S, 2020MNRAS.499..230B}, while more massive systems primarily rely on internal mechanisms \citep{2012MNRAS.424..232W, 2016MNRAS.457.4360Z}. Nevertheless, the primary drivers of galaxy evolution and their respective timescales remain open to debate.

Various models address the evolution of galaxies within clusters, from which the ``slow-then-rapid'' quenching model has been commonly adopted in the recent literature \citep{2013MNRAS.432..336W}. According to this model, galaxies infalling into clusters may initially experience a delay in the quenching of their star formation activity. \citet{2019ApJ...873...42R} suggest this delay phase results from a threshold in intra-cluster medium (ICM) density required to initiate environmental effects, particularly RPS. \citet{2022MNRAS.509..567S} propose that during the delay phase, galaxies undergo morphological transitions from spirals to lenticulars. Once the ICM density threshold is surpassed, RPS becomes highly efficient at removing the gas component of infalling galaxies, transitioning them from star-forming to a quiescent state.

The transition from star-forming to quiescence is evident as a bimodality in color-color diagrams \citep[e.g.,][]{2005ApJ...624L..81L, 2007A&A...476..137A, 2016A&A...586A..23D, 2016A&A...590A.102M}. These diagrams define three regions: 1) the blue cloud (BC) -- comprising mainly late-type star-forming galaxies; 2) the red sequence (RS) -- consisting of early-type galaxies with minimal star formation; and 3) the green valley (GV) -- an intermediate region with partially quenched galaxies of intermediate morphologies \citep{2009MNRAS.396..818S, 2010MNRAS.404..792M, 2010MNRAS.405..783M}. Although originally defined in color-color diagrams, the definition of BC, GV, and RS can be extend to the star formation main sequence (SFMS) diagram \citep{2014MNRAS.440..889S, 2020MNRAS.491.5406T, 2022MNRAS.509..567S}, which tracks the relation between stellar mass and star formation rate, regardless of environment. Various studies \citep{2014MNRAS.440..889S, 2017MNRAS.464.1077W, 2022MNRAS.509.3889D} suggest that different quenching mechanisms result in distinct paths from the BC to the RS \citep{2014MNRAS.440..889S, 2017MNRAS.464.1077W}. However, the SFMS cannot assess timescales, as it does not estimate the duration of the transition.

In the context of galaxies infalling into clusters, alongside star formation rate evolution, a dynamic evolution within the cluster is expected \citep{2011MNRAS.416.2882M, 2013MNRAS.431.2307O}. Recent infallers are typically found in highly radial orbits near the cluster outskirts, whereas virialized galaxies exhibit predominantly circular orbits within the cluster's inner region (virialized region). This allows connecting time evolution with the transition from radial to circular orbits using Phase Space \citep{2017ApJ...843..128R}. Due to projection effects along the line of sight, simulations are employed \citep{2016MNRAS.463.3083O, 2017ApJ...843..128R, 2019MNRAS.484.1702P} to correlate the time since infall ($t_{\rm inf}$) with location in a projected Phase Space (PPS). Various studies utilize PPS analysis, such as \citet{2019MNRAS.484.1702P}, that demonstrate that pre-processed galaxies exhibit distinct properties compared to pristine infalling galaxies, \citet{2020ApJS..247...45R} estimate the relationship between SFR and $t_{\rm inf}$, \citet{2020MNRAS.495..554R} observe RPS in the Coma Cluster, and \citet{2013MNRAS.434..784R, 2017MNRAS.467.3268R, 2019MNRAS.487L..86D, 2021MNRAS.503.3065S} show that different cluster dynamical stages result in distinct galaxy property distributions.

Combining the SFMS diagram with PPS and a substantial sample of galaxies provides insights into key questions regarding star formation quenching and morphological transitions for galaxies infalling into clusters. The Sloan Digital Sky Survey \citep[SDSS,][]{2000AJ....120.1579Y} offers an optimal dataset due to its extensive sky coverage and completeness, allowing the identification of several high-density (cluster) regions compared to galaxies evolving in isolation. By combining SDSS data with the aforementioned diagrams, we can quantify the temporal aspects of these processes and analyze their relation to the ``delayed-then-rapid'' quenching model. In this paper, we present a comparative analysis of the timescales for star formation suppression and morphological transitions, aiming to provide a new perspective on the seemingly conflicting results in the literature \citep{2009ApJ...707..250M, 2019MNRAS.486..868K, 2022MNRAS.509..567S}. Finally, we employ the BC and RS definitions to offer a dynamic estimation of the time galaxies spend in the GV after cluster infall and compare it to previous estimates \citep[e.g.,][]{2021MNRAS.508..157M}.

We organize this paper as follows: in \S2, we detail our selection criteria, present the cluster and field samples, and describe the sources of the galaxy properties we analyze. In \S3, we discuss the distribution of cluster and field galaxies in the SFMS diagram and in a morphological analogue, incorporating a morphological tracer on the y-axis. By combining these, we connect the transition from BC to RS with the morphological transition. In \S4, we explore the PPS to provide evidence that the evolution from BC to RS corresponds to an increase in the average $t_{\rm inf}$. We also investigate the distribution of BC, GV, and RS fractions as functions of infall time, using this information to estimate the GV timescale. In \S5, we examine the distribution of quenched and early-type galaxies in the PPS as functions of $t_{\rm inf}$, and provide insights into how morphological transition may differ for galaxies of varying stellar masses. In \S6, we present our main results and conclusions. Throughout this paper, we adopt a flat $\Lambda$CDM cosmology with $[\Omega_{M}, \Omega_{\Lambda}, H_{0}] = [0.27, 0.73, 72 \, {\rm km \, s^{-1} \, Mpc^{-1}}]$ to be consistent with \cite{2017AJ....154...96D} and report the magnitudes in the AB system.

\section{Data selection}

In this study, we use data from the Sloan Digital Sky Survey - 16th Data Release (SDSS-DR16) \citep{2020ApJS..249....3A}. Our sample was restricted to galaxies with redshifts in the range $0.03 \leq z \leq 0.1$ and Petrosian apparent magnitudes in the r-band ($m_{\rm r}$) brighter than 17.78, aligning with the survey's spectroscopic limit at $z = 0.1$. This minimum redshift constraint was applied to minimize biases in galaxy properties due to the fixed 3 arc-second fiber employed in the SDSS Legacy survey.

\subsection{Characterizing Galactic Environment}

Our primary objective is to compare the effects of high- and low-density environments on galaxy evolution. To achieve this, we utilize the Yang catalogue \citep{2007ApJ...671..153Y, 2008ApJ...676..248Y, 2009ApJ...695..900Y} for classifying galaxies based on their environmental characteristics. This catalogue is constructed by applying a halo mass finder algorithm to the New York University - Value Added catalogue \citep{2005AJ....129.2562B}. In this framework, groups are defined as galaxies residing within the same dark matter halo. The catalogue also distinguishes between central galaxies (the most massive in a cluster) and satellite galaxies, enabling the identification of isolated centrals, cluster/group centrals, and satellite galaxies.

The Yang catalogue, however, has been proven considerably restricted regarding satellite galaxies, which are important for studying the outer regions of clusters. Thus, we employ an alternative version of the Yang catalogue from \cite{2017AJ....154...96D}. In this version, galaxies are selected around the centers defined by the original catalogue, with membership assigned through an iterative shiftgapper technique \citep{2009MNRAS.399.2201L}. This technique has two main advantages: it does not assume the dynamical state of the cluster, and it extends membership selection to larger radii ($\sim 2.5R_{200}$) compared to the original definition ($\sim 1R_{200}$). Dynamical quantities are derived using Virial analysis (see the Appendix of \citealt{2021MNRAS.503.3065S}). Comparisons between the Halo Mass Finder and shiftgapper methods show differences smaller than 0.1 dex in estimated halo masses.

By selecting clusters with at least 20 members, we define a sample of 319 systems. The accuracy of cluster-centric coordinates decreases for clusters with fewer members due to Poissonian uncertainty. To ensure completeness in halo mass, we use the relation between $M_{200}$ and $N_{\rm members}$ to set a lower halo mass limit of $10^{14}M_{\odot}$. This reduces our sample to 254 clusters, comprising 20,191 galaxies, which we refer to as the ``cluster sample."

We also define a secondary sample of ``field galaxies"\footnote{Here ``field'' does not exclude galaxies in filaments \citep[e.g.][]{2014MNRAS.438.3465T}, since our focus is on understanding the role of the cluster environment.} to compare with the observed trends for clusters. To minimize the effects of pair interactions, we select isolated centrals from the Yang catalogue that are located beyond $5R_{180}$ from any structure with halo masses greater than $10^{13}M_{\odot}$. The radius $R_{180}$ is calculated using the following scaling relation \citep{2009ApJ...695..900Y}:
\begin{equation}
R_{180} \sim 1.61 , \left( \frac{M_{\rm halo}}{10^{14}{\rm M_{\odot}}} \right)^{1/3} , (1+z_{\rm group})^{-1} \rm Mpc.
\end{equation}
Additionally, to avoid the effects of survey edges, we include only galaxies located within the SDSS main follow-up area ($120 \leq \rm RA \leq 250$, $0 \leq \rm DEC \leq 60$).

Furthermore, a galaxy being the sole resident of its halo does not guarantee that its evolution was dominated by internal mechanisms. For example, fossil groups, which result from a sequence of merging events, ultimately form a single, highly luminous central galaxy. As shown by \cite{2009AJ....137.3942L}, fossil groups are characterized by the presence of a bright galaxy with an absolute magnitude $M_{\rm r} \leq -22$ in the r-band. Therefore, we use this magnitude as the lower limit for our field sample to mitigate the influence of fossil groups. After these selections, our secondary sample consists of 11,674 field galaxies.

\subsection{Estimating galaxy properties}

\subsubsection{MPA-JHU catalogue}

We use stellar mass\footnote{Derived using the methodology of \cite{2003MNRAS.341...33K}}, star formation rate, and specific star formation rate estimates\footnote{Estimates are calculated using the \cite{2004MNRAS.351.1151B} prescription and then aperture corrected using the method described in \cite{2007ApJS..173..267S}} from the Max-Planck Institute für Astrophysik - John Hopkins University (MPA-JHU) catalogue. This catalogue provides measurements of absorption and emission line properties for galaxies in the SDSS-DR16 database that do not exhibit anomalies in their spectra. Notably, 98\% of the galaxies in our combined sample (cluster and field) have valid estimates.

\subsubsection{T--Type: a continuous way to define morphology}

In this work, we utilize the T--Type parameter, first introduced by \cite{1963ApJS....8...31D}, aimed at identifying S0 galaxies, which have T--Type $\sim 0$. We select a revised version of T--Type as presented in \cite{2018MNRAS.476.3661D}. This parameter is derived continuously by employing equation (7) from \cite{2015MNRAS.446.3943M}:

\begin{equation}
\text{T--Type} = -4.6P(\text{Ell}) -2.4P(\text{S0}) + 2.5P(\text{Sab}) + 6.1P(\text{Scd})
\end{equation}

where $P({\rm X})$ denotes the probability of a galaxy being classified as a given morphology, with X representing Elliptical (Ell), lenticular (S0), A-B spiral (Sab), and C-D spiral (Scd). \cite{2018MNRAS.476.3661D} derived the probabilities for different morphological types by applying a deep-learning Convolutional Neural Network (CNN) algorithm calibrated using responses from the Galaxy Zoo 2 \citep{2013MNRAS.435.2835W, 2016MNRAS.461.3663H} survey as input. The models demonstrated high accuracy ($\geq 97\%$) when applied to a test sample with characteristics similar to those used for calibration. The final catalogue provides T--Type estimates for a subsample of approximately 670,722 galaxies from the SDSS-DR7 database, which encompasses both our cluster and field samples. Due to the high precision in T--Type estimate, we do not expect uncertainty greater than unity in T--Type. Therefore, hereon, we adopt ${\rm T\text{--}Type} = 0.5$ and $-0.5$ to distinguish spiral galaxies from S0 galaxies, and S0 galaxies from elliptical galaxies, respectively.

\subsection{Assessing Sample Completeness}

\subsubsection{Mass completeness}

\begin{figure}
    \centering
    \includegraphics[width = 0.8\columnwidth]{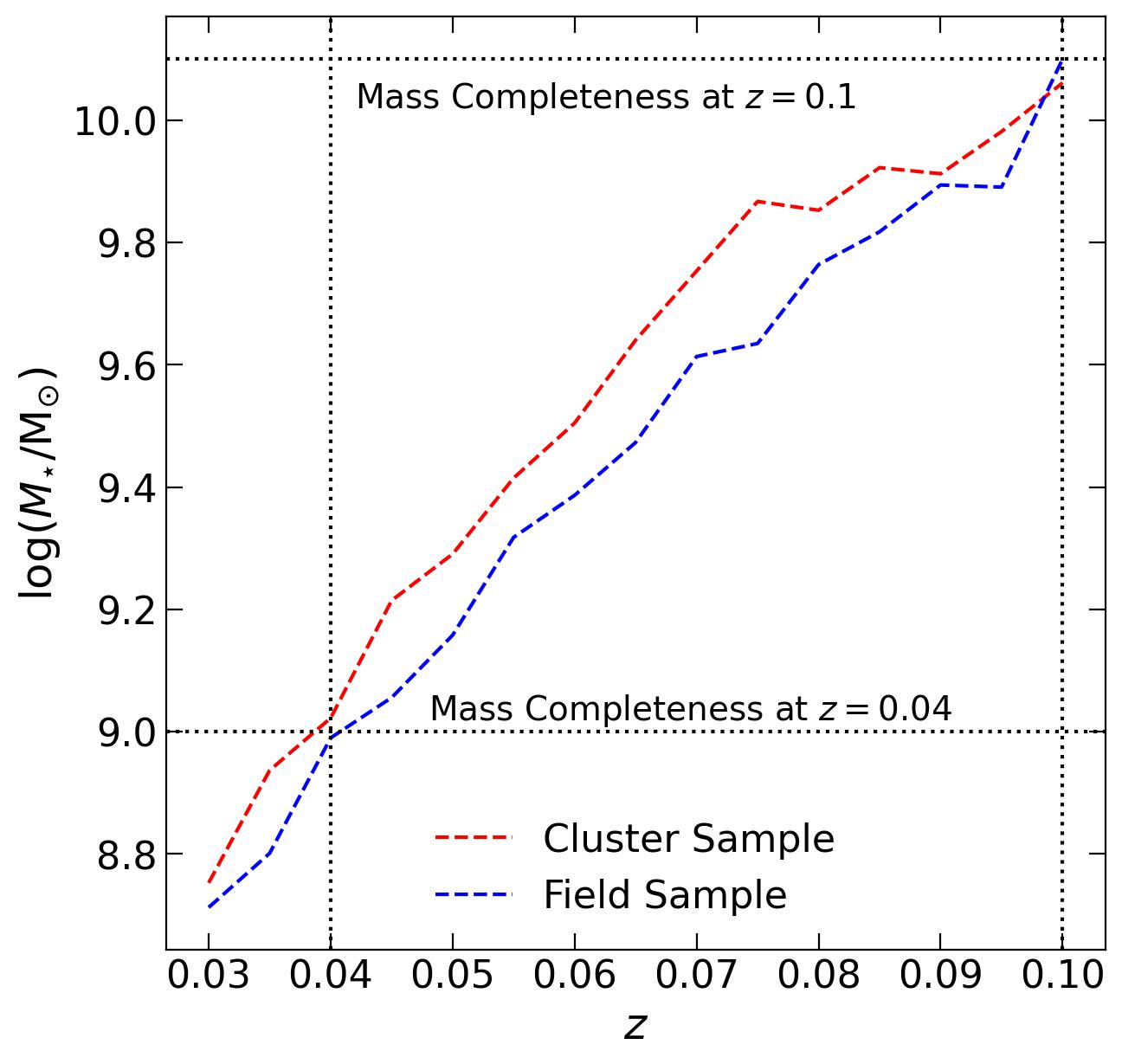}
    \caption{Stellar mass completeness as a function of redshift for the cluster (red curve) and field (blue) samples. The black dashed lines denote the mass completeness at $z=0.1$ (top line) and $z = 0.04$ (bottom line). In the following, when considering galaxies below $10^{10}M_{\odot}$, we restrict our analysis to $z \leq 0.04$.}
    \label{fig:mass_completeness}
\end{figure}

To guarantee that our sample is representative of the physical processes affecting galaxies' evolution, it is pivotal to assess the mass completeness as a function of redshift. We divide galaxies into redshift bins with a width of 0.005 and compute the 95\% quantile in the stellar mass distribution for each bin. Given that the cluster and field samples have different selection functions, this procedure is performed separately for both. In Fig.~\ref{fig:mass_completeness} we show our mass completeness as a function of redshift for both cluster (red) and field (blue) samples. Notably, there is an average offset of approximately 0.1 dex between the cluster and field samples.

In Figure~\ref{fig:mass_completeness}, the upper black dashed line indicates that our dataset is complete across the entire redshift range for galaxies with stellar masses greater than $10^{10.1} \rm{M_{\odot}}$, regardless of environmental factors. To ensure completeness for galaxies with stellar masses down to $10^{9} \rm{M_{\odot}}$, we establish a secondary redshift limit. The lower black dashed line sets this threshold at $z=0.04$. Consequently, we analyze the full redshift range for galaxies with stellar masses exceeding $10^{10}  \rm{M_{\odot}}$, while our analysis for galaxies with stellar masses between $10^{9} \rm{M_{\odot}}$ and $10^{10} \rm{M_{\odot}}$ is restricted to the redshift range $0.03 \leq z \leq 0.04$. Although different redshift ranges are applied to different stellar mass bins, we anticipate minimal evolution in the properties of galaxies within the studied range ($0.03 \leq z \leq 0.1$).

\subsubsection{Completeness at small radii}
\label{sec:compl_radii}

\begin{figure}
    \centering
    \includegraphics[width = 0.8\columnwidth]{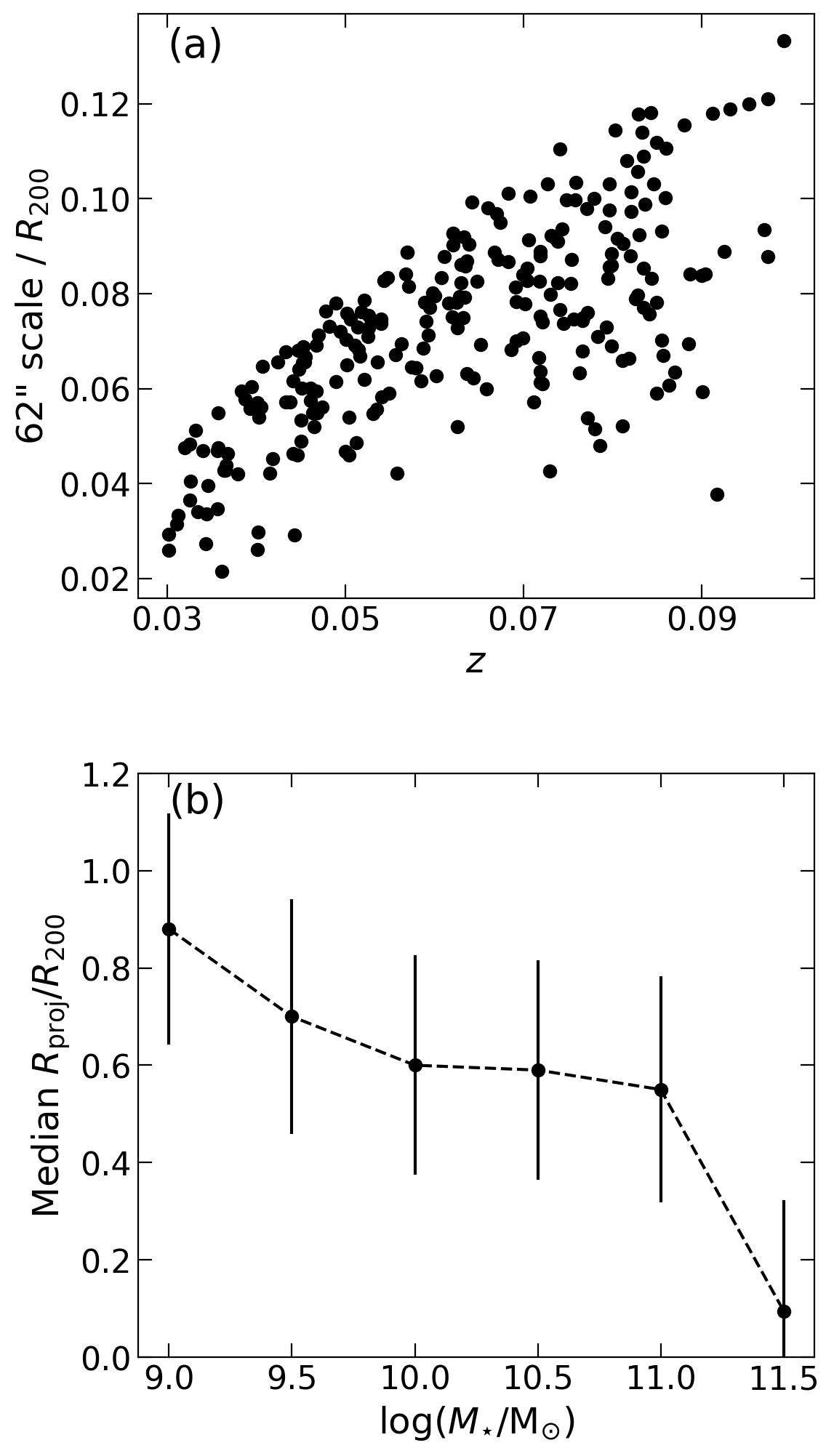}
    \caption{Top: ratio between the fiber collision limiting 62'' scale and $R_{200}$ for clusters at different redshifts. Bottom: median $R_{\rm proj}/R_{200}$ as a function of stellar mass. We limit our analysis to the $R_{\rm proj}/R_{200} \geq 0.1$ region in clusters (see text).}
    \label{fig:radii_completeness}
\end{figure}

In particular, for our cluster sample, the SDSS Legacy survey suffers from fiber collision effects, such that no two fibers can be placed closer than 62 arc-seconds. This limitation means that galaxies closer than a certain threshold, depending on redshift, cannot both be observed. To quantify this effect, we consider the ratio between the 62'' distance converted to kpc at each redshift and the $R_{200}$ for each cluster. Panel (a) of Figure~\ref{fig:radii_completeness} shows the distribution of this ratio as a function of redshift for the clusters in our sample. On average, the 62'' scale represents $7 \pm 2 \%$ of $R_{200}$.

In addition to the instrument limitations, the intracluster light in the inner region of clusters is dominated by the brightest cluster galaxy (BCG), which can hinder the detection of faint, low-mass objects in the vicinity of the BCG. To address this issue, we investigate the median $R_{\rm proj}/R_{200}$ as a function of stellar mass. The results are shown in panel (b) of Figure~\ref{fig:radii_completeness} and suggest that the most massive galaxies primarily dominate the region with $R_{\rm proj}/R_{200} \leq 0.1$. This observation aligns with the principle of mass segregation \citep{1980ApJ...241..521C}, where massive galaxies are predominantly located in the cores of clusters, while less massive systems are found in the outskirts.

To minimize both instrumental and contamination from massive galaxies' light in our conclusions, we hereon limit our analysis to the $R_{\rm proj}/R_{200} \geq 0.1$ region of clusters.

\section{Parameter space of galaxy evolution}

\begin{figure}
    \centering
    \includegraphics[width = \columnwidth]{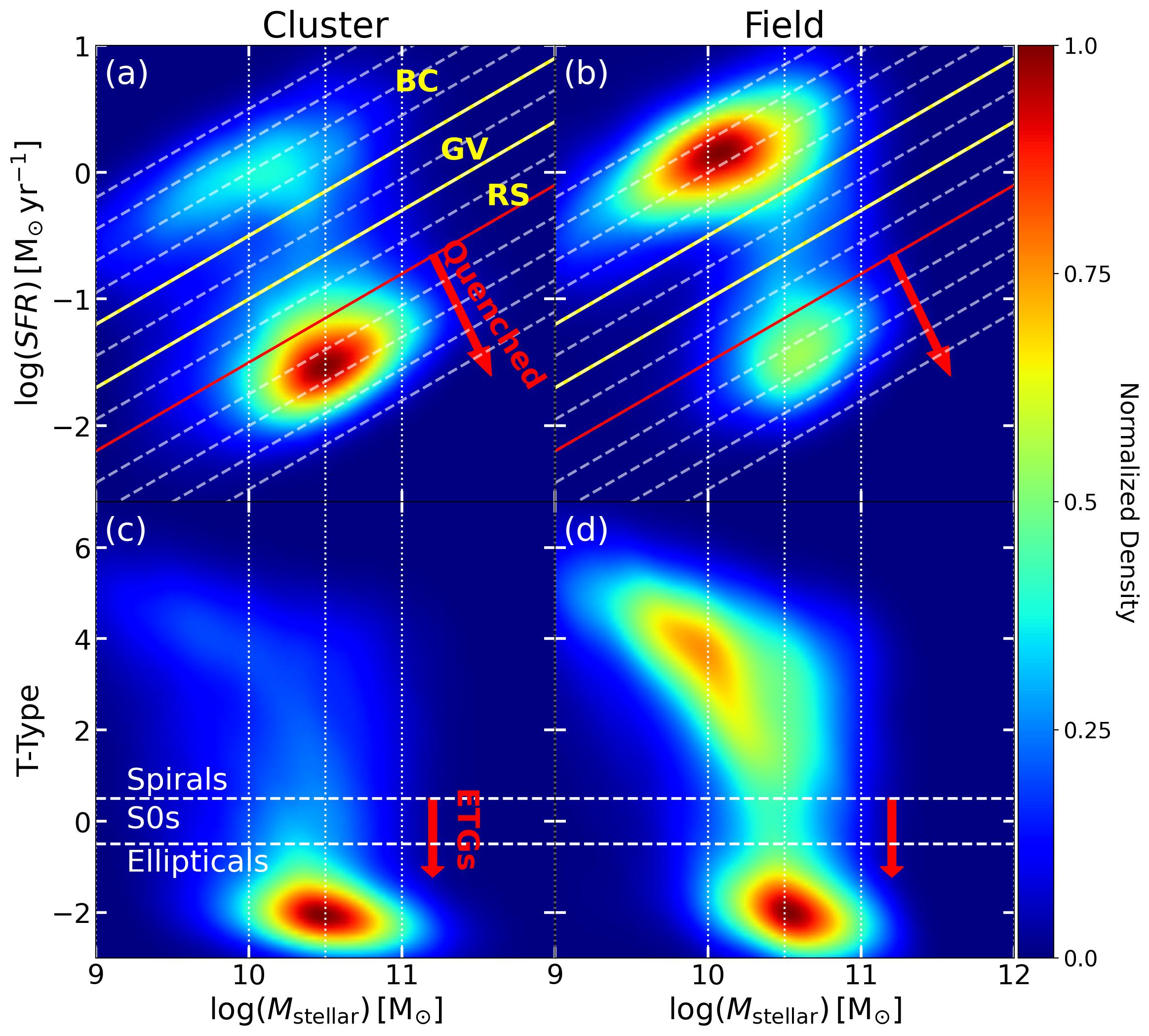}
    \caption{Smoothed distributions in the SFMS (top row) and T--Type diagram (bottom row) for cluster (left column) and field (right) samples. The smoothing is done using a kernel density estimator with width equal to 1\% of the X- and Y-axis ranges. In the SFMS diagram we also show the limiting lines between BC, GV and RS in yellow, slices with the same slope but varying intercept as the SFMS as white dashed lines, and a red line and arrow denoting what we select as ``quenched galaxies". In the T--Type diagram we include the threshold lines between Spiral, S0 and Elliptical morphologies, alongside a red arrow denoting what we call ``early-type'' galaxies.}
    \label{fig:SFR_TType_Distribution}
\end{figure}

Galaxies can be characterized within a parameter space defined by observable properties, aiding in the understanding of their evolutionary stages. The parameters considered in this study include stellar mass, star formation rate (SFR), and morphology (T--Type). In Fig.~\ref{fig:SFR_TType_Distribution}, we present the distribution of cluster (left column) and field (right column) samples in the SFMS diagram and T--Type versus Stellar Mass (bottom row). These distributions were smoothed using a Gaussian kernel density estimator with a width corresponding to 1\% of the ranges of the X and Y axes. In the SFMS panels, the yellow dividing lines BC/GV and GV/RS are shown, determined respectively by the equations \citep{2020MNRAS.491.5406T, 2022MNRAS.509..567S}:
\begin{equation}
    \log(SFR/{\rm M}_{\odot} {\rm yr}^{-1}) = 0.7 \log(M_{\star}/{\rm M_{\odot}}) - 7.5
\end{equation}
\begin{equation}
    \log(SFR/{\rm M}_{\odot} {\rm yr}^{-1}) = 0.7 \log(M_{\star}/{\rm M_{\odot}}) - 8.
\end{equation}
For the T--Type diagram, we indicate the separation spiral/S0 and S0/elliptical as white dotted lines in the bottom panels of Fig.~\ref{fig:SFR_TType_Distribution}. In all panels, vertical lines denote the boundaries of the four adopted stellar mass bins. Henceforth, we define ``quenched'' and ''early-type'' galaxies as those below the thresholds indicated by the red arrows or lines in the relevant panels.

Examining the SFMS diagram, we observe that cluster galaxies predominantly occupy a single high-density region within the red sequence (RS), whereas field galaxies exhibit two distinct regions: a primary one in the blue cloud (BC) and a secondary, less pronounced one in the RS. This disparity between field and cluster galaxies suggests environmental effects are driving star-forming galaxies towards a quiescent state. Notably, this effect is primarily observed in galaxies with stellar masses $\log(M_{\star}/\rm M_{\odot}) < 10$. Above this mass threshold, the differences between cluster and field galaxies diminish, indicating that more massive galaxies are less susceptible to environmental influences and rely predominantly on internal mechanisms to suppress star formation. Consequently, we define stellar mass bins to distinguish populations whose evolution is governed by different mechanisms.

When examining morphology, we find that the comparison between cluster and field galaxies yields results consistent with those observed in the SFMS analysis. Specifically, there is a notable similarity in the stellar mass range where peak densities are found for elliptical galaxies and RS galaxies, pointing to an overabundance of massive, quenched, elliptical galaxies in clusters. In contrast, field galaxies exhibit a greater diversity of star-forming systems, reflected in a distribution extending towards higher T--Types. This implies that star formation in field environments occurs across a broader range of spiral morphologies.

\begin{figure}
    \centering
    \includegraphics[width = \columnwidth]{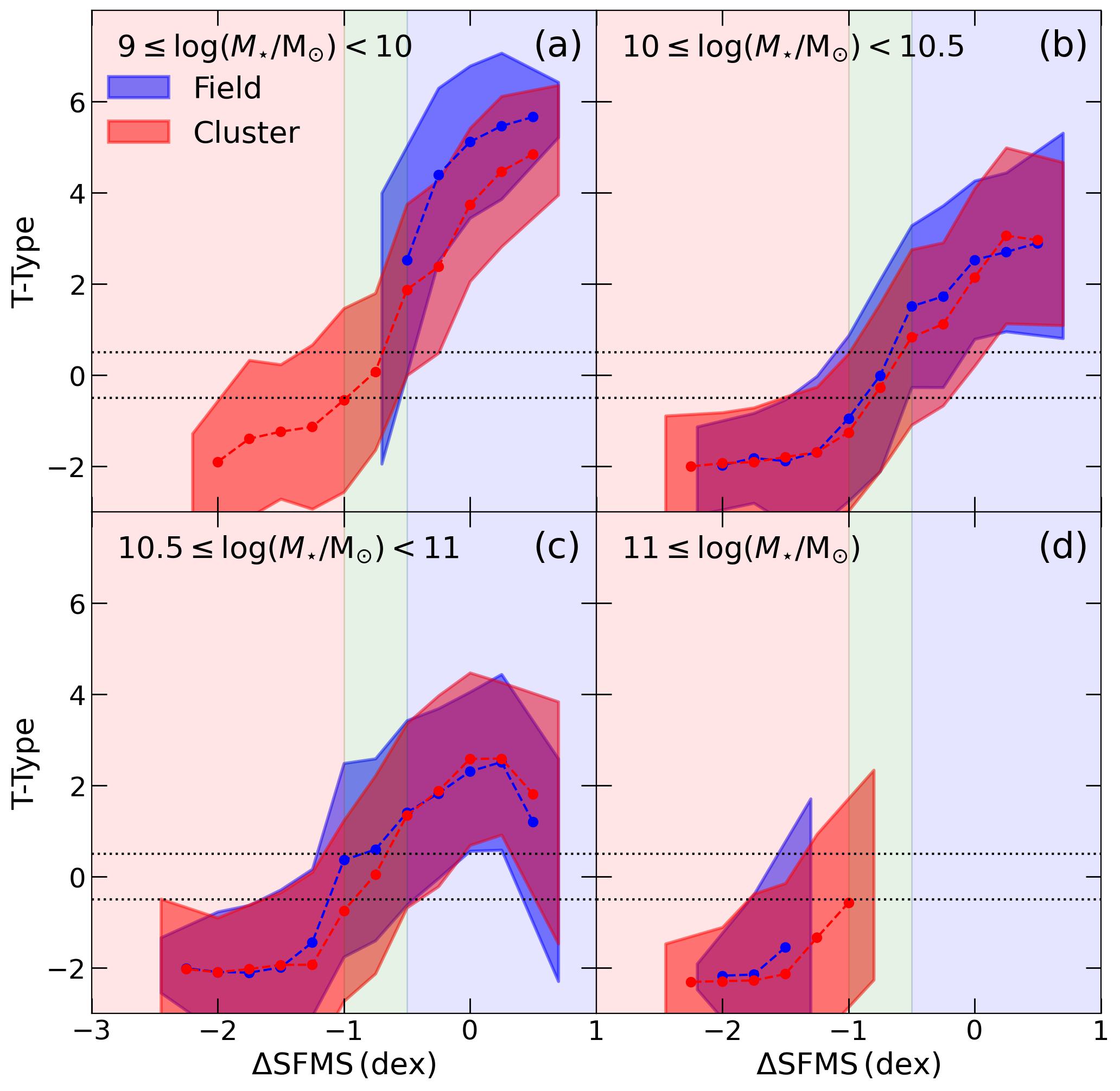}
    \caption{T--Type as a function of the distance from the SFMS for cluster (red) and field (blue) samples, separating galaxies according to their stellar mass. The circles joined by solid lines represent the median of the distributions, while the shaded areas correspond to the observed 1-$\sigma$ scatter. The background colors denote the BC, GV and RS regions and the horizontal lines are the morphological thresholds presented in Fig.~\ref{fig:SFR_TType_Distribution}.}
    \label{fig:TType_deltaSFMS_evolution}
\end{figure}

The broad range of morphologies exhibited by star-forming systems, contrasted with the relatively narrow range observed in quenched systems, underscores the non-linear relationship between morphology and star formation rate. While the SFMS diagram clearly delineates star-forming from quenching or quenched galaxies across different stellar masses, the correlation between morphology and stellar mass in the T--Type diagram prevents such distinct separation. Critically, the dependence of morphology on stellar mass suggests that ``morphological transition'' may entail different variations in T--Type across various stellar mass bins. For instance, it is posited that high-mass elliptical galaxies are remnants of major merger events in the early universe. Additionally, it has been proposed that spirals, even those undergoing suppression of star formation, may retain their stellar disks and evolve into S0s \citep{1996hst..prop.6480D, 1997hsth.conf..185D}.

Hence, it is crucial to establish a method for comparing star formation rate and morphology effectively. We utilize the large number of galaxies in our sample to segment the SFMS diagram into more than three discrete regions, facilitating the tracking of T--Type variations from the BC to the RS. The regions we analyze are delineated by white dashed lines in panels (a) and (b) of Fig.~\ref{fig:SFR_TType_Distribution} and are described by the equation
\begin{equation}
\log({\mathrm{SFR} \times \mathrm{yr}^{-1} \mathrm{M}_{\odot}}) = 0.7\log(M_{\star}/{\rm M_{\odot}}) - i,    
\end{equation}
with $i$ varying from 6.5 to 9.5 in steps of 0.25. We further define the Star Formation Main Sequence (SFMS) as the best linear fit to all galaxies (cluster and field) within the BC. This unified SFMS ensures that subsequent analyses reference the same baseline for both cluster and field galaxies, particularly when describing the distance from the SFMS. The SFMS zero-point is determined to be
\begin{equation}
    \log({\mathrm{SFR} \times \mathrm{yr}^{-1} \mathrm{M}_{\odot}}) = (0.7 \pm 0.05) \log(M_{\star}/{\rm M_{\odot}}) - (7 \pm 0.2).
\end{equation}

In Fig.~\ref{fig:TType_deltaSFMS_evolution}, we display the median (dashed lines) and 1-sigma scatter (shaded regions) of the T--Type distribution as a function of the perpendicular distance from the SFMS ($\Delta \mathrm{SFMS}$), segregated by environment (red for cluster galaxies and blue for field galaxies) and divided into four stellar mass bins (each panel). It is noteworthy that testing different numbers of bins revealed no significant changes in the observed trends. The background colors represent the BC, GV, and RS regions as defined in Fig.~\ref{fig:SFR_TType_Distribution}. Horizontal lines indicate the T--Type thresholds for Spiral, S0, and Elliptical galaxies.

Comparison between cluster and field galaxies reveals decreasing differences in T--Type with increasing stellar mass. The average differences are $\Delta \mathrm{T{-}Type}(\mathrm{Cluster{-}Field}) = -0.83 \pm 0.05$, $-0.16 \pm 0.03$, $0.09 \pm 0.03$, $0.02 \pm 0.03$\footnote{Errors denote the statistical error in the mean}, respectively. Low-mass systems are more susceptible to losing components through ram pressure stripping (RPS) and/or tidal mass loss. However, in the highest mass bin, the minimal differences between cluster and field galaxies may suggest that these systems have experienced similar evolutionary paths. Nonetheless, it is important to consider that we are analyzing a low-redshift sample, implying that most of the massive systems had undergone significant evolution at higher redshifts, as predicted by the downsizing scenario. Consequently, drawing definitive conclusions about the physical mechanisms influencing these galaxies is challenging. For instance, due to the high-density environment, massive galaxies in the cores of clusters may have experienced a higher merger rate in early epochs compared to their field counterparts.

More importantly, it is noticeable that galaxies emerge from the green valley (GV) with $\text{T--Type} \leq 0$, irrespective of stellar mass and environment. This consistent trend suggests that star formation suppression is intertwined with the fading and/or removal of prominent star-forming disks in spiral galaxies. This observation aligns with the reported excess of S0 galaxies in the GV \citep{2018MNRAS.477.4116K, 2022MNRAS.509..567S}. Additionally, it indicates that galaxies reach early-type morphologies while there is still room for a gradual decrease in the star formation rate towards the bottom of the RS. Although we adopt $\text{T--Type} \sim 0$ to define galaxies that have undergone morphological transition, our conclusion remains valid even if we use a more restrictive $\text{T--Type} = -2$ as the morphological transition boundary.

These results suggest that the fading and/or suppression of spiral arms occurs prior to a galaxy reaching its final quenched state, with this removal being more likely for low-mass galaxies in clusters, at least in the local universe. However, when investigating the SFMS diagram, time evolution is not directly measurable, as we are limited to observing the present state of a given galaxy. Therefore, to derive time-scales related to the transition between different regions, we invoke the dynamical evolution expected for galaxies within clusters, viewed through the perspective of Projected Phase Space.

\begin{figure}
    \centering
    \includegraphics[width = \columnwidth]{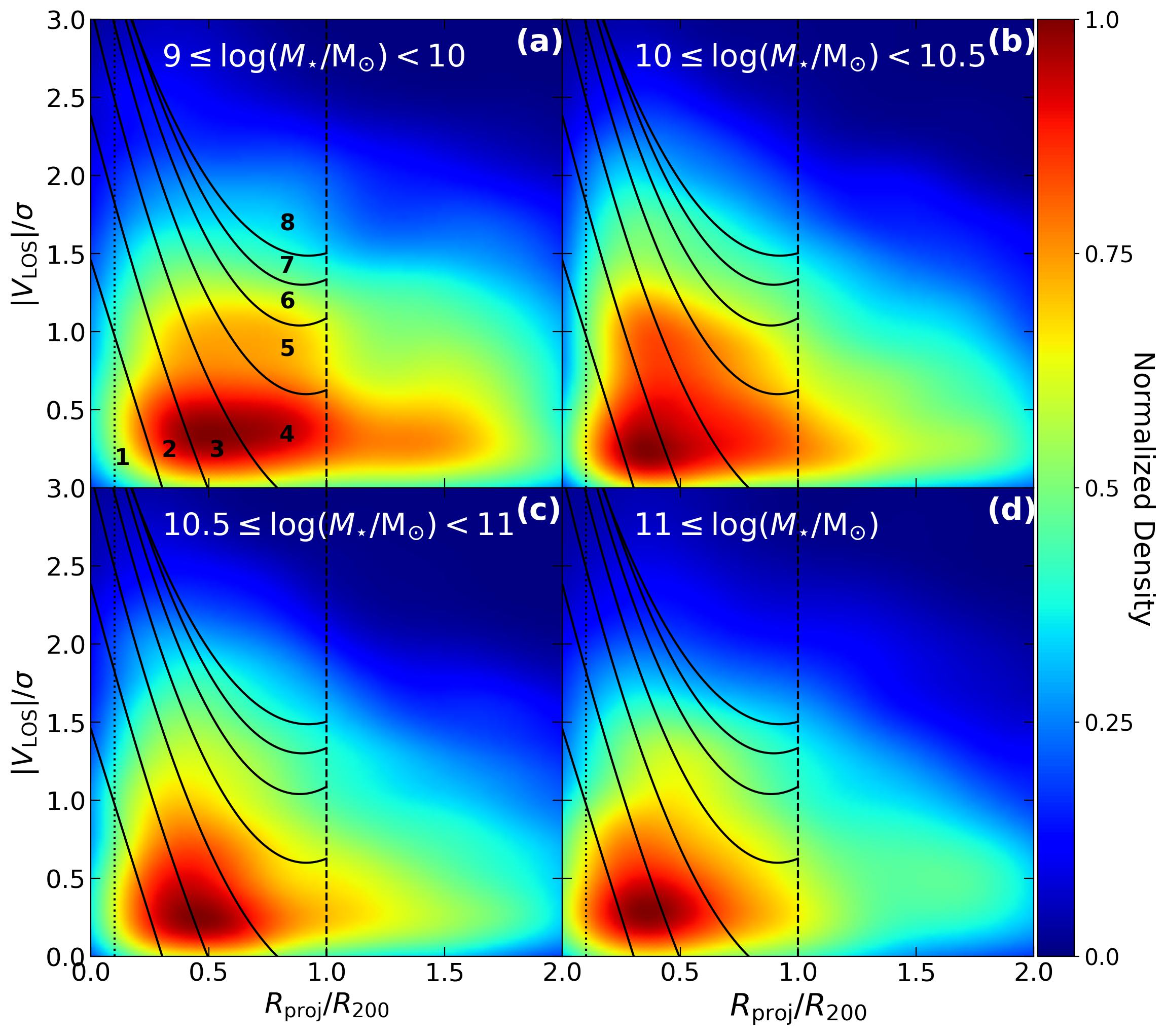}
    \caption{Normalized density distribution of galaxies in the PPS. We separate galaxies into four different stellar mass bins. The solid black lines show the different regions we analyse and relate to a given infall time \citep{2019MNRAS.484.1702P}. These regions are limited to within $R_{200}$, which is shown as a vertical dotted line.}
    \label{fig:PPS_distribution}
\end{figure}

\section{The projected phase space}

\begin{figure}
    \centering
    \includegraphics[width = \columnwidth]{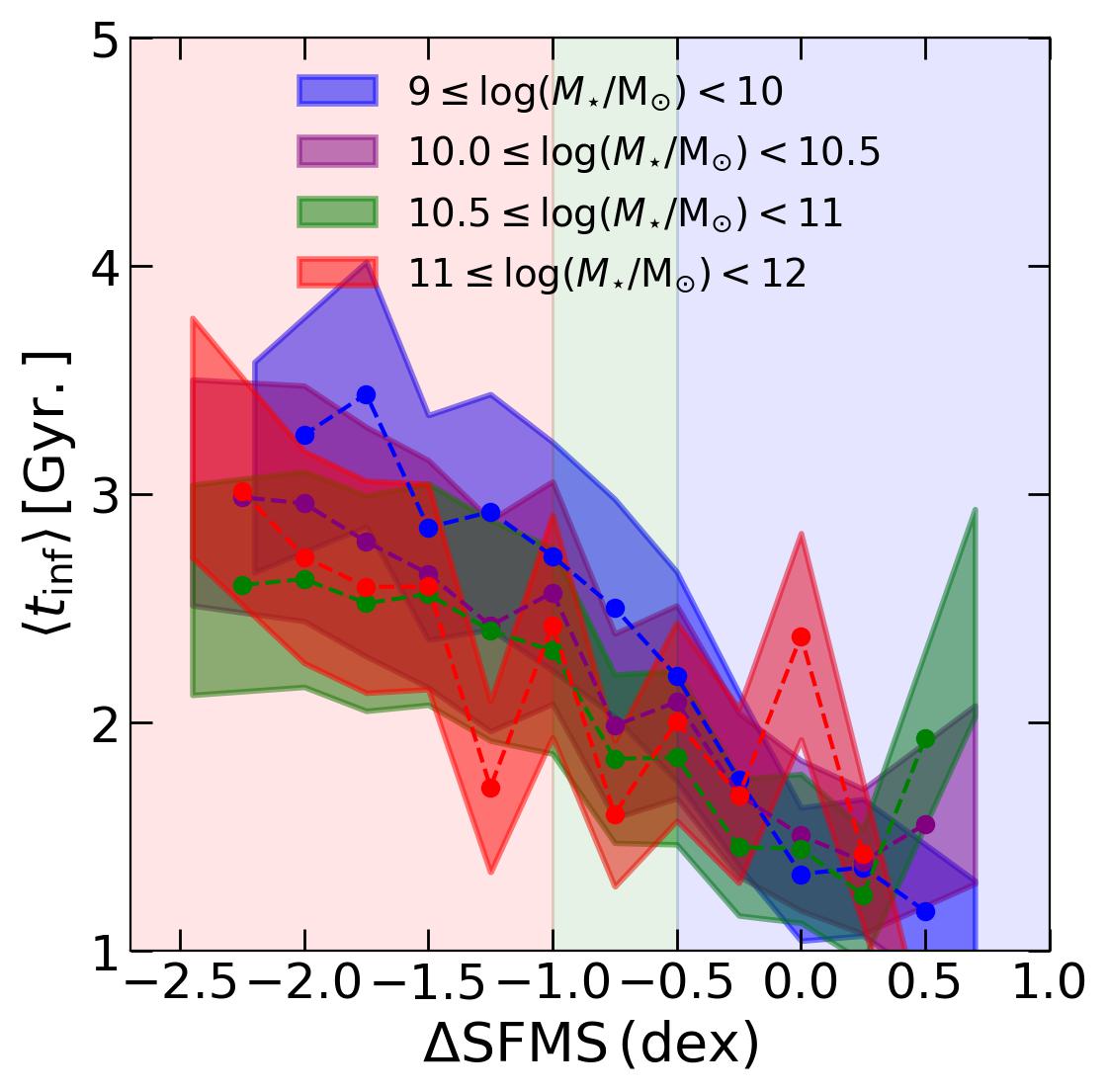}
    \caption{Average infall time as a function of $\Delta \rm SFMS$ for four different stellar mass bins. The circles joined by solid lines represent the median of the distributions, while the shaded areas correspond to the observed 1-$\sigma$ scatter. The background colors represent the BC, GV and RS regions.}
    \label{fig:tinf_vs_deltaSFMS}
\end{figure}

The evolution of galaxies infalling into clusters can be understood through their trajectories in Phase Space. This allows us to establish a relationship between the time since infall ($ t_{\rm inf}$), defined as the time since a galaxy first crossed the $R_{200}$ radius, and its location in Phase Space. Despite limitations and additional uncertainties, simulations also enable average infall times to be associated with different regions of the Projected Phase Space (PPS), where only line-of-sight velocities and projected sky positions are available. Here, we use the slicing method presented in \citet{2019MNRAS.484.1702P}. Utilizing the YZiCS zoom-in simulation \citep{2017ApJ...837...68C}, \citet{2019MNRAS.484.1702P} investigated the time-since-infall distribution in the projected phase space and defined regions that constrain galaxies to a narrow time-since-infall range. These regions are constructed for $R_{\rm proj} \leq R_{200}$, but for completeness, we also qualitatively examine galaxy properties beyond this threshold. Additionally, we tested our results with two alternative methods of dividing the PPS: (1) \citet{2017ApJ...843..128R}, which assigns a probability of a galaxy having a given time-since-infall within each region; and (2) \citet{2020ApJS..247...45R}, which employs a square grid in the PPS. We find no significant differences in the results across these methods. In Fig.~\ref{fig:PPS_distribution}, we present the distribution of galaxies in the PPS. We categorize galaxies into different stellar mass bins and smooth the distribution using a Gaussian kernel density estimator with a width equal to 1\% of the X- and Y-axis ranges. The solid black lines delineate regions containing galaxies with different $t_{\rm inf}$ values: 1.42, 2.24, 2.77, 3.36, 3.89, 4.50, 5.18, and 5.42 Gyr, respectively, for regions 8 down to 1 \citep{2019MNRAS.484.1702P}. The dotted line indicates the lower radial threshold adopted to avoid contamination by very massive (bright) galaxies in the cluster core (see Section \ref{sec:compl_radii}).

Comparison between the different stellar mass bins shows that the high-density envelope (red/orange colors) of low-mass galaxies (panel a) extends beyond $R_{200}$, whereas most of the massive systems are primarily constrained within $R_{200}$. This spatial distribution is consistent with the concept of mass segregation \citep{1980ApJ...241..521C} and is in agreement a dynamical structure divided into two components: a virialized core, dominated by the most massive, red, and dead galaxies; and an outer semi-equilibrium region where most of the infalling and/or backsplash galaxies are found \citep{1997ApJ...476L...7C, 2001ApJ...547..609E}.

\begin{figure*}
    \centering
    \includegraphics[width = \textwidth]{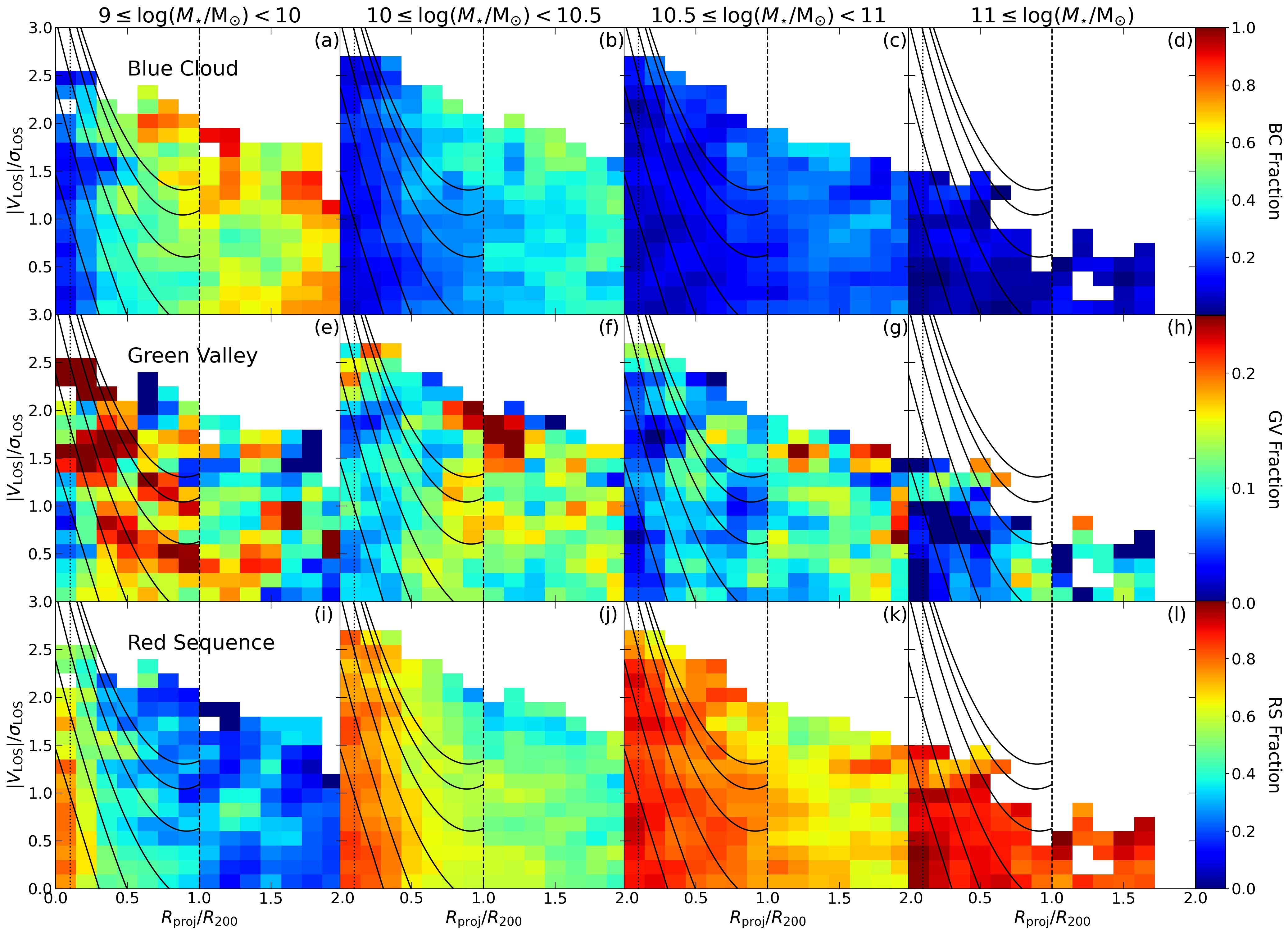}
    \caption{Distribution of (from top to bottom) BC, GV and RS galaxies in the Projected Phase Space. Galaxies are divided according to stellar mass. The solid black lines define values of constant infall time. The vertical dotted line denotes the $R_{200}$ threshold, which is the upper limit in radii in which the defined regions are valid. The color-bar range is selected to highlight relevant trends. We highlight that the fraction of galaxies in the green valley is considerably smaller to the observed fraction range observed in the BC and the RS.}
    \label{fig:pps_bc_gv_rs}
\end{figure*}

Our analysis relies on the connection between time since infall and the evolutionary time sequence of galaxies. However, galaxies in region 8 are not necessarily the progenitors of those observed in region 1. Nevertheless, our upper time since infall is 5.42 Gyr, which corresponds to $z \sim 0.5$. Galaxies formed most of their stellar mass ($\sim 75\%$) within $0.5 \leq z \leq 2.5$ \citep{2014ARA&A..52..415M}, indicating that we do not expect significant variations in the mass assembly history of galaxies within the analyzed infall time range. Additionally, the variation in the SFMS from $z \sim 0.5$ to $z \sim 0.1$ is $\leq 0.3$ dex \citep[e.g.][]{2014ApJ...795..104W, 2016ApJ...817..118T, 2022MNRAS.509.5382C}, suggesting small variations in the properties of galaxies being accreted onto clusters over the analyzed time-since-infall range. Therefore, as a first approximation, we can consider the time-since-infall as an evolutionary sequence for infalling galaxies.

To demonstrate the correspondence between the evolution in the SFMS diagram and the PPS, we present Fig.~\ref{fig:tinf_vs_deltaSFMS}, which illustrates the relationship between $t_{\rm inf}$ and $\Delta \rm SFMS$ across four distinct stellar mass bins. Similar to Fig.~\ref{fig:TType_deltaSFMS_evolution}, the background colors indicate the BC, GV, and RS regions. It is noticeable that $t_{\rm inf}$ increases as $\Delta \rm SFMS$ decreases, indicating progression from the BC to the RS. The minor variations across different stellar mass bins are consistent with findings from \citet{2019MNRAS.484.1702P}, suggesting that projection effects blend galaxies with varying infall times. Notably, this relationship offers a rough estimate of the time galaxies spend in the GV, which appears to be approximately 0.8 Gyr. However, the inherent uncertainties in $t_{\rm inf}$ limit our analysis to a qualitative assessment. Thus, we infer that the GV represents a relatively brief phase in a galaxy's evolutionary timeline, compared to the BC and RS.

\subsection{Evolution from blue cloud to red sequence}
\begin{figure*}
    \centering
    \includegraphics[width = \textwidth]{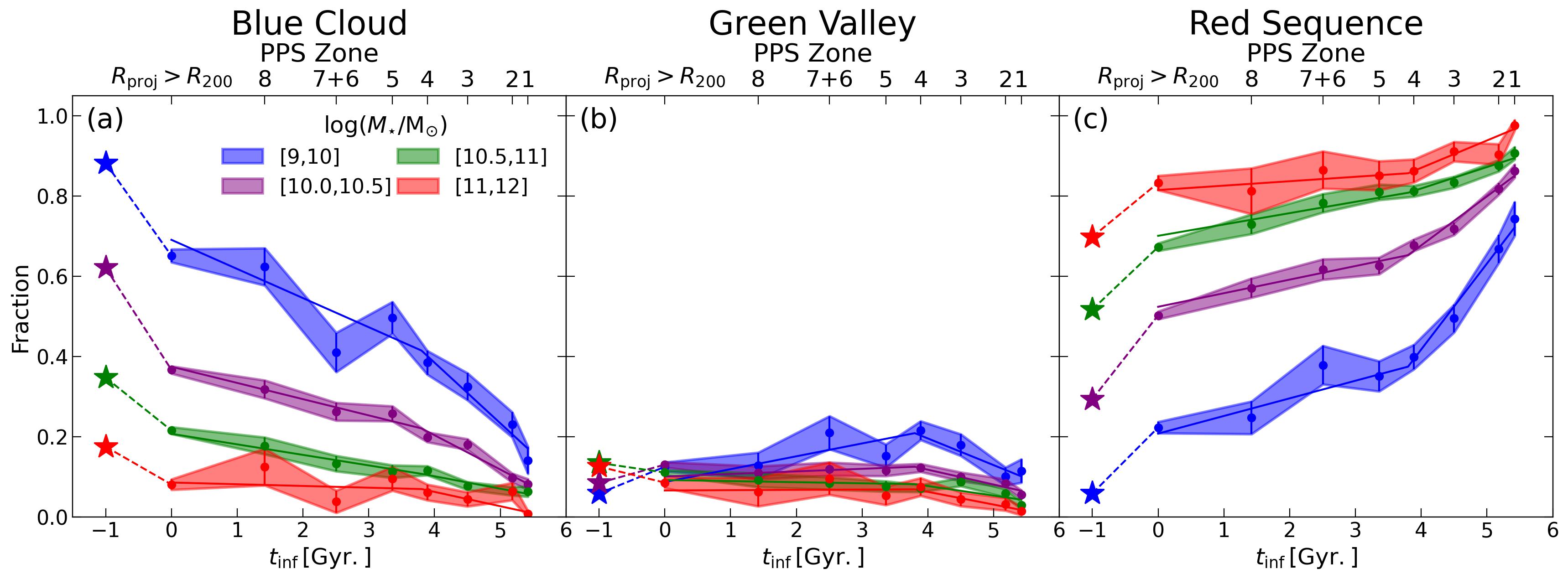}
    \caption{Relation between fraction of galaxies in the BC (left), GV (center) and RS (right) as a function of infall time. For each stellar mass bin, the error-bars and shaded areas the median and 1-sigma scatter. For completeness, we also include two extra points in the relations: 1) the field counterpart value, symbolized with a star of the respective colour (at $ t_{\rm inf} = -1$ Gyr.); and 2) the fraction of BC, GV and RS galaxies beyond $R_{200}$ (at $ t_{\rm inf} = 0$ Gyr.). The double linear model fitted in each curve is shown as solid lines.}
    \label{fig:tinf_bc_gv_rs}
\end{figure*}
Our initial investigation examines the distribution of BC, GV, and RS galaxies within the Projected Phase Space (PPS). In Fig.~\ref{fig:pps_bc_gv_rs}, we present these distributions, segmented by stellar mass. We only include pixels containing at least 20 galaxies to guarantee robustness, and adjust the color-bar range to emphasize relevant trends. It is noteworthy that the fraction of green valley galaxies is consistently smaller than that of the BC and RS galaxies, regardless of stellar mass. However, distinct trends in the green valley distribution fractions are observed across different mass bins. In panel (e), a relatively high object density is found between regions 3 to 7 (as labeled in Fig.~\ref{fig:PPS_distribution}), corresponding to intermediate infall times and suggesting environmental influences on these galaxies. This peak, indicative of intermediate times since infall, is statistically significant and is observed exclusively in the lower stellar mass bin. Additionally, the absence of the most massive galaxies at higher radii and velocities is consistent with mass segregation.

In our analysis of galaxy fractions within the Projected Phase Space (PPS), we discern an inverse symmetry between the distributions of BC and RS galaxies across all stellar masses. This symmetry is primarily due to the relatively constant fraction of green valley (GV) galaxies, with only slight variations observed throughout the PPS. A particularly notable trend is the marked increase in the fraction of RS galaxies beyond the $R_{200}$ threshold during the transition from BC to RS. These results suggest two potential scenarios: first, the influence of the cluster environment on galaxy evolution may extend to larger radii, likely driven by harassment or tidal effects \citep{2017MNRAS.467.3268R}; second, this observation may reflect the peak in the fraction of backsplash galaxies within the 1-2 $R_{200}$ range \citep{2011MNRAS.416.2882M, 2020MNRAS.492.6074H, 2023MNRAS.519.4884F}. As galaxies move beyond $R_{200}$, the decline in BC galaxy fractions appears to align with regions characterized by varying times since infall, thereby suggesting a significant link between the infall process and subsequent galaxy evolution.

Another relevant aspect of our findings is the pronounced reduction in the fraction of BC galaxies, which is concurrently accompanied by an increase in the fraction of RS galaxies. This transition becomes particularly noticeable within the inner regions of the cluster, specifically at approximately $\sim 0.3R_{200}$. At this radial distance, the fraction of BC galaxies drops below 20\%, while the fraction of RS galaxies rises above 50\%. This observation aligns with the expectation that environmental mechanisms, such as ram pressure stripping, become increasingly effective at removing galactic components as galaxies move closer to the cluster core. The correlation between this radial threshold and the steep increase in environmental density further supports this scenario \citep{2019ApJ...873...42R}.

\subsection{Addressing the Slow-then-Rapid Quenching model}

In this section, we focus in analyzing the fraction of BC, GV, and RS galaxies as a function of time-since infall, in order to assess galaxy evolution over time. Figure \ref{fig:tinf_bc_gv_rs} illustrates the variation in the fractions of BC (panel a), GV (panel b), and RS (panel c) galaxies as a function of infall time for different stellar mass bins, each represented by distinct colors. To provide a comprehensive overview, we introduce two supplementary points in each panel: 1) the fraction in the field sample, denoted by the same color star symbol at $t_{\rm inf} = -1$; and 2) the fraction at larger radii ($R_{\rm proj} > R_{200}$), marked at $t_{\rm inf} = 0$\footnote{This infall time is manually set, acknowledging that the outer regions of clusters may host a significant fraction of back-splash galaxies \citep{2013MNRAS.431.2307O, 2023MNRAS.519.4884F}.}. The x-axis coordinates for these two points are chosen for illustrative purposes and will not be employed in any quantitative analyses.

\begin{table*}
\caption{Double line model fitting results for the different curves shown in Fig.~\ref{fig:tinf_bc_gv_rs}. $a_{1}$ and b denote the slope and intercept, respectively, of the line fitting the $ t_{\rm inf} < t_{\rm delay}$ region. $a_{2}$ denotes the slope of the line fitting the $ t_{\rm inf} > t_{\rm delay}$. Both $a_{1}$ and $a_{2}$ are given in the units of $\rm Gyr^{-1}$. We divide the sample according to the SFMS region and into different stellar mass bins.}
\label{tab:bc_gv_rs_model}
\resizebox{\textwidth}{!}{%
\begin{tabular}{c|ccc|ccc|ccc}
\hline
\multirow{2}{*}{$\log(M_*/{\rm M}_\odot)$} & \multicolumn{3}{c|}{Blue cloud}                       & \multicolumn{3}{c|}{Green valley}                    & \multicolumn{3}{c}{Red sequence}                    \\ \cline{2-10} 
                                           & $a_1$            & $b$             & $a_2$            & $a_1$           & $b$             & $a_2$            & $a_1$           & $b$             & $a_2$           \\ \hline
{[}9,10{]}                                 & $-0.07 \pm 0.01$ & $0.70 \pm 0.03$ & $-0.15 \pm 0.03$ & $0.04 \pm 0.03$ & $0.07 \pm 0.03$ & $-0.06 \pm 0.02$ & $0.04 \pm 0.03$ & $0.21 \pm 0.01$ & $0.21 \pm 0.04$ \\
{[}10,10.5{]}                              & $-0.04 \pm 0.01$ & $0.38 \pm 0.04$ & $-0.09 \pm 0.02$ & $0.01 \pm 0.02$ & $0.10 \pm 0.04$ & $-0.03 \pm 0.03$ & $0.03 \pm 0.02$ & $0.52 \pm 0.03$ & $0.12 \pm 0.02$ \\
{[}10.5,11{]}                              & $-0.02 \pm 0.02$ & $0.20 \pm 0.04$ & $-0.03 \pm 0.02$ & $0.00 \pm 0.01$ & $0.09 \pm 0.04$ & $-0.02 \pm 0.02$ & $0.03 \pm 0.02$ & $0.70 \pm 0.02$ & $0.05 \pm 0.03$ \\
{[}11,12{]}                                & $0.00 \pm 0.02$  & $0.08 \pm 0.05$ & $-0.03 \pm 0.02$ & $0.00 \pm 0.02$ & $0.06 \pm 0.02$ & $-0.01 \pm 0.01$ & $0.01 \pm 0.01$ & $0.81 \pm 0.01$ & $0.07 \pm 0.02$ \\ \hline
\end{tabular}
}
\end{table*}
Utilizing Fig.~\ref{fig:tinf_bc_gv_rs}, we provide a quantitative view of the trends observed in Fig.~\ref{fig:pps_bc_gv_rs}. We highlight the symmetry between BC and RS fractions, coupled with a GV fraction that remains relatively constant over time. The distinctions between the fractions of BC and RS galaxies in both the field and at small infall times ($t_{\rm inf} < 2$ Gyr) tend to amplify with decreasing stellar mass, conforming to expectations where BC galaxies are more prevalent in the field, while RS galaxies dominate in clusters. Notably, the GV fraction remains approximately constant across all infall times, albeit with some mass dependence: for low-mass galaxies, there is a discernible increase in the fraction of GV galaxies with intermediate $t_{\rm inf}$, suggesting that for these galaxies, the environment plays a role in quenching their star formation.

Significantly, the behavior of BC and RS fractions relative to infall time reveals a consistent trend across all stellar mass bins. We observe two distinct phases: initially, before a critical time $t_{\rm delay}$, the BC and RS fractions exhibit minimal variation. However, post $t_{\rm delay}$, there is a sharp decline in the BC fraction, mirrored by an increase in the RS fraction. This transition is strongly dependent on stellar mass, with the most pronounced changes occurring in the lower mass bins.

Our extended analysis of the PPS beyond $R_{200}$ supports a two-step quenching process. Initially, as galaxies cross $R_{200}$, they encounter minimal environmental influences, with tidal mass loss emerging as the likely mechanism during this stage \citep{2017ApJ...843..128R, 2019MNRAS.490..773R}. This tidal interaction may reduce the stellar mass of galaxies, potentially causing a shift to lower mass bins compared to their progenitors. This phase is characterized by slow transformation. Subsequently, a more rapid transformation phase occurs around $\sim 0.6R_{200}$ (approximately $\sim R_{500}$), where the denser intracluster medium (ICM) facilitates strong ram pressure stripping (RPS), effectively removing the majority of the galaxy's gas component. This two-step process aligns with the slow-then-rapid quenching model \citep[e.g.][]{2012MNRAS.424..232W}, and we can pinpoint the infall time at which the transition from the slow to the rapid quenching phase occurs.

\subsection{The in-between: green valley time-scale}

The results shown in Fig.~\ref{fig:tinf_bc_gv_rs} allow us to estimate the timescale galaxies spend in the green valley (GV). The relatively constant GV fraction indicates a balance between the inflow and outflow of galaxies within this region. It is important to note that, although galaxies can undergo rejuvenation \citep[e.g.][]{2010ApJ...714L.171T, 2015Galax...3..192M, 2019MNRAS.489.1265M}, most of their mass is already assembled in the adopted redshift range. Furthermore, the cluster environment is expected to predominantly drive the quenching of star formation, making episodes of rejuvenation rare. Thus, galaxy evolution can be statistically approximated as a unidirectional transition from the BC to the RS. Consequently, we can utilize the BC and RS fractions to estimate the duration of environmental effects required to populate the GV with the observed fractions.

We model the curves as a function of time since infall using a double line profile:
\begin{equation}
\centering
\label{eq:double_line}
F(t_{\rm inf}) = 
     \begin{cases}
       a_{1}t_{\rm inf} + b, & \text{if} \, \, t_{\rm inf} < t_{\rm delay} \\
       a_{2}t_{\rm inf} + (a_{1} - a_{2})t_{\rm delay} + b, & \text{if} \,\, t_{\rm inf} \geq t_{\rm delay},
     \end{cases}
\end{equation}
where the second and third terms in the $t_{\rm inf} \geq t_{\rm delay}$ line ensure continuity at $t_{\rm inf} = t_{\rm delay}$. It is important to note that our dataset contains a limited number of data points, and allowing all variables to be free parameters may result in overfitting. To address this, we reduce the degrees of freedom by fixing $t_{\rm delay} = 3.8$ Gyr. This value is chosen based on fitting procedures conducted for both the BC and RS fraction relations, with the average of the resulting $t_{\rm delay}$ distribution used. The model results are presented in Table \ref{tab:bc_gv_rs_model} and are presented as solid lines in Fig.~\ref{fig:tinf_bc_gv_rs}.

\begin{figure*}
    \centering
    \includegraphics[width = \textwidth]{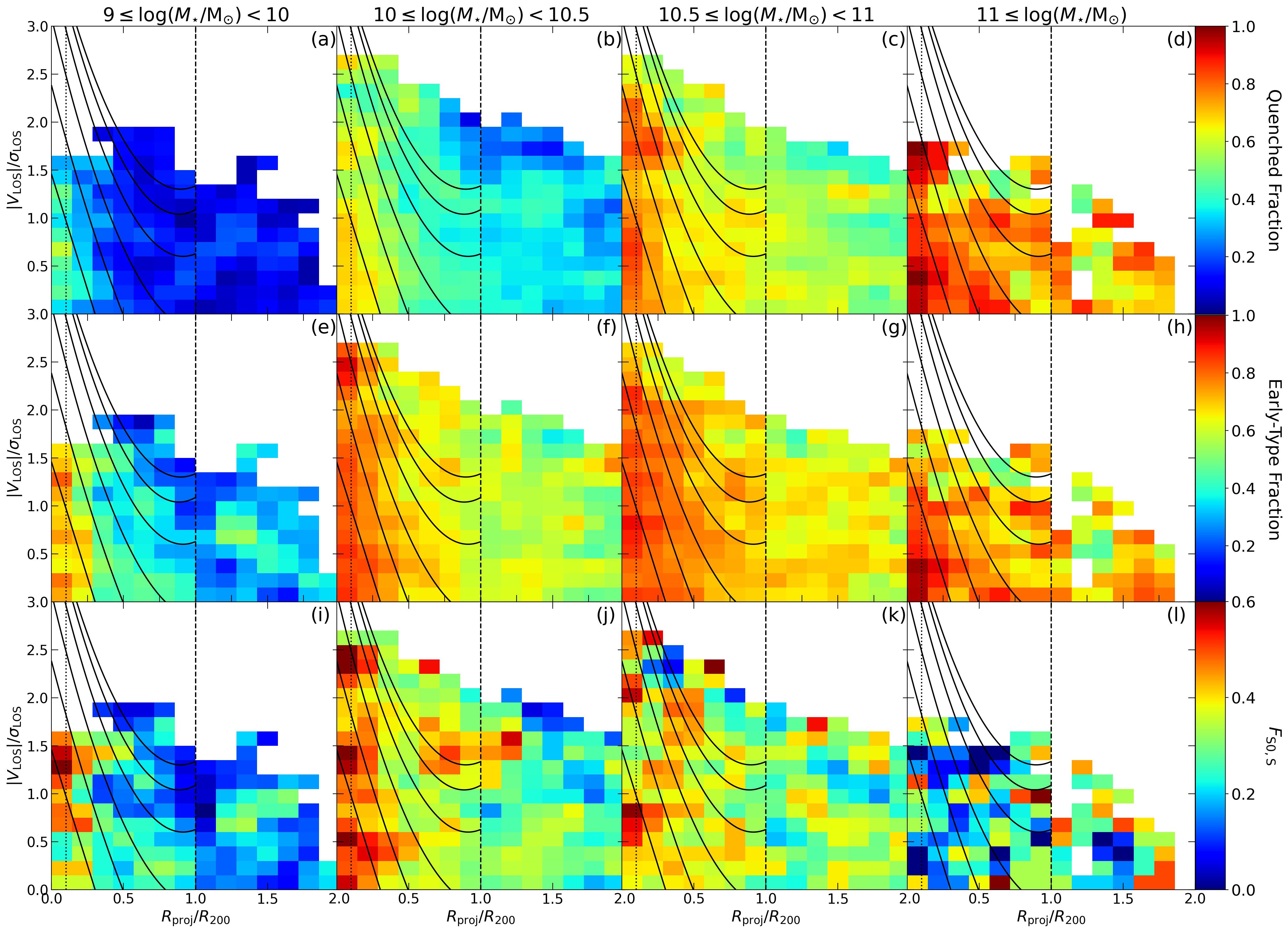}
    \caption{The distribution of quenched (top row) and early-type (bottom row) fraction in the PPS. Each column represent a different stellar mass bin. The thresholds for calling a galaxy quenched and/or early-type are shown in Fig.~\ref{fig:SFR_TType_Distribution}. The solid black lines show the different regions comprising galaxies with a narrow range of time since infall. For calculating the fractions in a given pixel we impose a minimum number of 20 galaxies to guarantee robustness.}
    \label{fig:Quenched_Fraction_PPS}
\end{figure*}

The slope for the $t_{\rm inf} < t_{\rm delay}$ region is approximately zero, irrespective of galaxy mass and the region considered (BC, GV, or RS). For the intercept $b$, we observe a decreasing trend with increasing stellar mass for BC galaxies, and an increasing trend for RS galaxies, as anticipated. In the GV, the intercept value remains nearly constant across different masses, barely exceeding the 10\% mark. For $t_{\rm inf} \geq t_{\rm delay}$, we observe an increase in the steepness of the BC and RS relations regardless of stellar mass, whilst the slope modelling the GV remains virtually zero. This supports the hypothesis of a balance between the inflow and outflow of galaxies in this region.

The observed trends allow us to estimate the time galaxies spend in the GV ($\Delta t_{\rm inf}^{\rm GV}$) by combining the $a_{2}$ slopes of the BC and RS relations:
\begin{equation}
    F^{\rm GV} = \Delta t_{\rm inf}^{\rm GV} (a_{2}^{\rm BC} - a_{2}^{\rm RS}),
\end{equation}
where $F^{\rm GV}$ denotes the average fraction of galaxies in the GV. Our analysis focuses on separating environmental effects from the internal evolution of galaxies, highlighting that the derived time-scale mainly reflects environmentally-driven quenching processes. Concentrating on the two lower stellar mass bins, which are most affected by the environment, we find that galaxies would cross the GV due to environmental effects in approximately $0.5$ and $1.1$ Gyr, respectively, for the two lower stellar mass bins. This indicates a shorter time-scale for galaxies with lower stellar mass. Our estimates are consistent (within 1-sigma uncertainty) with time-scales derived for clusters at $1 \leq z \leq 1.4$ \citep{2021MNRAS.508..157M}. However, the trend we observe with mass in significantly different from their work. While we find a decreasing time-scale with stellar mass, their estimate remain roughly constant. This may follow from several possibilities: 1) large uncertainty in their estimate at higher masses; 2) adopted different stellar mass bins; and 3) at higher redshifts, more massive galaxies were actively forming stars, such that the variation caused by the environment is more significant in comparison to the local universe, where it seems to affect only low stellar mass systems.

\section{Comparing the timescales for star-formation suppression and morphological transformation}

We now compare the time-scales for star formation suppression and morphological transition. These processes are closely linked, and defining the boundaries for quenched galaxies and those that have undergone morphological transition is crucial. This analysis is intended to also consider the underlying morphology-density relation in galaxy clusters \citep{Dressler}.

\subsection{Insights from the distribution of quenched and early-type Galaxies}

In Fig.~\ref{fig:Quenched_Fraction_PPS}, we present PPS diagrams showing the fractions of quenched galaxies (first row) and early-type galaxies (second row) across different stellar mass bins (each column). Additionally, we display the ratio $F_{\rm S0,S} = N_{\rm S0}/(N_{\rm S0}+ N_{\rm S})$ in the last row, where $N_{\rm S0}$ represents the number of S0 galaxies ($-0.5 \leq\, \text{T--Type} \,\leq 0.5$) and $N_{\rm S}$ denotes the number of spiral galaxies ($0.5 \leq\, \text{T--Type}$). To ensure statistical robustness, each pixel includes at least 20 galaxies. Notably, for the highest stellar mass bin, galaxies primarily occupy the $\sim 0.5 R_{200}$ region, as shown in panels (d), (h), and (l). These trends support models suggesting that clusters have a virialized inner core predominantly composed of massive elliptical quenched galaxies.

Panels (a), (b), (c), and (d) reveal an increasing quenched fraction with stellar mass, consistent with the downsizing scenario. We observe that the quenched fraction reaches its maximum within the $R_{\rm proj} < 0.4 R_{200}$ range, regardless of stellar mass. Quantitatively, the average differences in quenched fractions between the $R_{\rm proj} < 0.4 R_{200}$ and $R_{\rm proj} > R_{200}$ regions are $19 \pm 3\%$, $23 \pm 5\%$, $16 \pm 6\%$, and $14 \pm 8\%$, respectively, for increasing stellar mass.

The early-type fraction, shown in panels (e), (f), (g), and (h), increases both with stellar mass and with decreasing $R_{\rm proj}$. These trends mirror those observed for quenched fractions, suggesting a link between star formation quenching and morphological transition. Notably, in the lowest stellar mass bin (panel e), there is a marked increase in the fraction of early-type galaxies after crossing $R_{200}$. Specifically, the average fraction beyond $R_{200}$ is $27 \pm 8\%$, compared to $40 \pm 7\%$ within $R_{\rm proj} < R_{200}$, indicating that $R_{200}$ may act as an initial threshold for morphological transition. However, this trend needs further investigation, as backsplash galaxies, which are primarily found near $R_{200}$, could influence this pattern.
\begin{figure*}
    \centering
    \includegraphics[width = \textwidth]{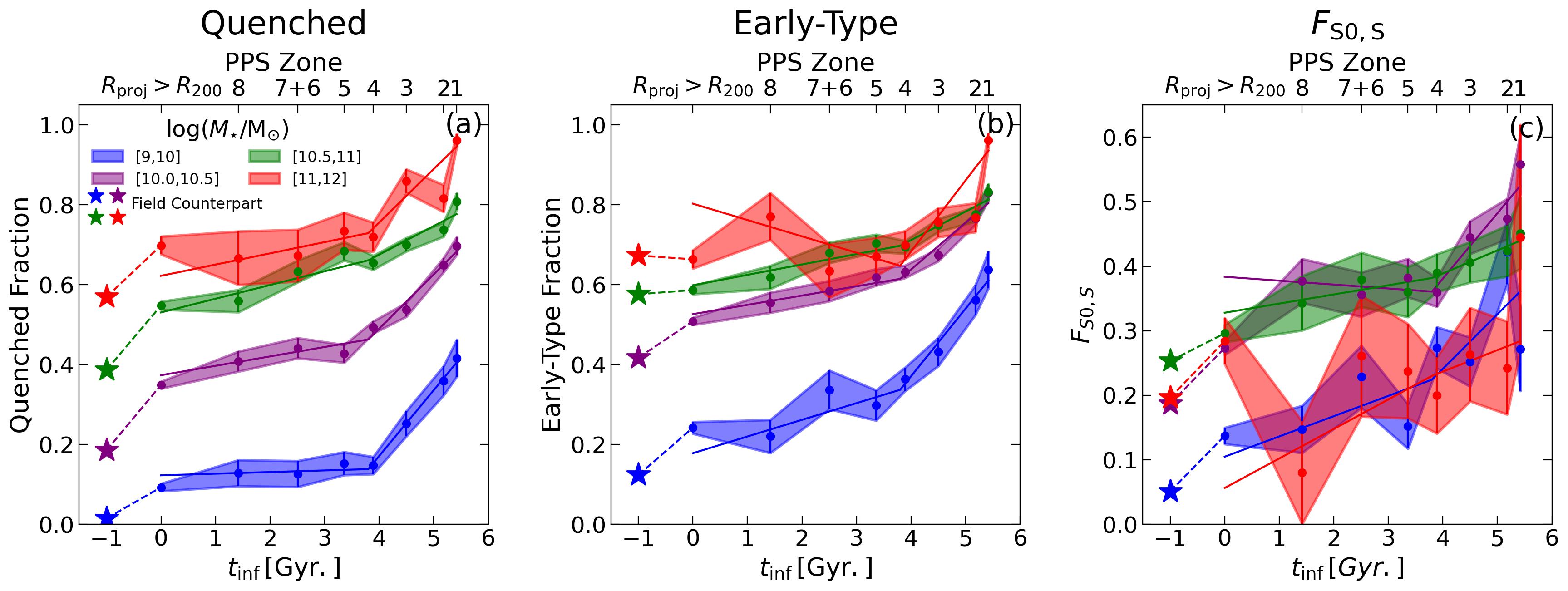}
    \caption{Similar to Fig.~\ref{fig:tinf_bc_gv_rs}, but showing the quenched (left) and early-type (center) fractions, and the $F_{\rm S0,S}$ ratio (right).}
    \label{fig:tinf_q_e_s0}
\end{figure*}

Panels (i) to (l) offer valuable insights into whether galaxies infalling into clusters evolve towards elliptical morphologies or retain a remnant disk, thus becoming S0s. By comparing the early-type fraction (panels e to h) with the $F_{\rm S0,S}$ ratio (panels i to l), we observe mass-dependent trends. For the two lower stellar mass bins, the trends in the early-type fraction align with those in the $F_{\rm S0,S}$ ratio, suggesting that the increase in early-type fraction is primarily due to a rise in the number of S0s. Conversely, in the most massive bins, a significant disparity between $F_{\rm S0,S}$ and the early-type fraction suggests that these massive early-type galaxies are predominantly ellipticals.

Our findings indicate a possible coupling between star formation quenching and morphological transition for galaxies of varying stellar mass across different PPS regions. From a qualitative analysis, we suggest that morphological transition may be initiated at larger radii compared to star formation quenching and is more pronounced with decreasing stellar mass. The similarities between early-type fraction and $F_{\rm S0,S}$ for low-mass galaxies suggest that these galaxies are primarily evolving into S0s. Conversely, for the most massive systems, the trends observed in the early-type fraction distributions do not align with those in the $F_{\rm S0,S}$ distribution. This disparity suggests that the morphological evolution leading to massive ellipticals differs significantly from the process experienced by spiral galaxies infalling into clusters. In the next section, we will quantitatively explore these differences and derive the associated time-scales.

\subsection{Quantifying environmental effects}

Fig.~\ref{fig:tinf_q_e_s0} illustrates the variation of quenched (panel a) and early-type (panel b) fractions, along with the $F_{\rm S0,S}$ ratio (panel c) as a function of infall time. The error bars and filled areas represent the median and 1-sigma scatter. We divide our sample according to stellar mass, and include $R_{\rm proj} > R_{200}$ and field estimates, similar to Fig.~\ref{fig:tinf_bc_gv_rs}. As in Fig.~\ref{fig:tinf_bc_gv_rs}, we begin our analysis by comparing field and cluster galaxies. The quenched fraction (panel a) comparison between cluster and field environments mirrors the trends observed in panel (c) of Fig.~\ref{fig:tinf_bc_gv_rs}. However, we observe a mass dependence in the comparison of early-type fractions between cluster and field environments. For the two lower stellar mass bins, there is a significant excess ($>10\%$) of early-type galaxies in clusters compared to their field counterparts. Conversely, this excess is only seen in the two most massive bins in the $t_{\rm inf} \geq 4$ Gyr region, whereas in the $t_{\rm inf} < 4$ Gyr region the early-type fraction in the field is similar to that observed in clusters. This enhanced fraction of early-type galaxies in the $t_{\rm inf} > 4$ Gyr region may result from massive galaxies being affected by the cluster environment only in inner regions, or it may reflect a pre-existing cluster structure where the most massive ellipticals are found in the core. This mass dependence highlights the role of the cluster environment in driving the morphological transition of low-mass galaxies. In contrast, massive systems either rely primarily on internal mechanisms to alter their morphology, or they enter the cluster environment after experiencing pre-processing effects in medium-density environments, such as small groups or filaments.

Exploring the relations with infall time, we find trends consistent with the slow-then-rapid quenching model. More importantly, the similar trends observed in both the quenched and early-type fractions suggest two major points: 1) the slow-then-rapid quenching model can be extended to include morphological variation; and 2) there may exist a causal relationship between quenching and morphological transition, where the removal of outer components of galaxies in clusters plays a crucial role in star formation suppression.

To quantify the observed trends, we adopt a similar double linear relation to the one presented in equation~\ref{eq:double_line}. The parameter fit results are presented in Table~\ref{tab:q_e_s0_model}. The fitted lines are also presented as solid lines in Fig.~\ref{fig:tinf_q_e_s0}.

\begin{table*}
\caption{Similar to Tab.~\ref{tab:bc_gv_rs_model}, but for curves shown in Fig.~\ref{fig:tinf_q_e_s0}. While $b$ is dimensionless, $a_{1}$ and $a_{2}$ are given in $\rm Gyr^{-1}$.}
\label{tab:q_e_s0_model}
\resizebox{\textwidth}{!}{%
\begin{tabular}{c|ccc|ccc|ccc}
\hline
\multirow{2}{*}{$\log(M_*/{\rm M}_\odot)$} & \multicolumn{3}{c|}{Quenched fraction}              & \multicolumn{3}{c|}{Early-type fraction}             & \multicolumn{3}{c}{$F_{\rm S0,S}$}                  \\ \cline{2-10} 
                                           & $a_1$           & $b$             & $a_2$           & $a_1$            & $b$             & $a_2$           & $a_1$           & $b$             & $a_2$           \\ \hline
{[}9,10{]}                                 & $0.00 \pm 0.01$ & $0.12 \pm 0.03$ & $0.16 \pm 0.03$ & $0.04 \pm 0.02$  & $0.18 \pm 0.01$ & $0.17 \pm 0.02$ & $0.03 \pm 0.02$ & $0.11 \pm 0.04$ & $0.11 \pm 0.03$ \\
{[}10,10.5{]}                              & $0.02 \pm 0.01$ & $0.38 \pm 0.04$ & $0.13 \pm 0.02$ & $0.02 \pm 0.02$  & $0.53 \pm 0.04$ & $0.11 \pm 0.03$ & $0.00 \pm 0.02$ & $0.38 \pm 0.05$ & $0.10 \pm 0.02$ \\
{[}10.5,11{]}                              & $0.03 \pm 0.01$ & $0.53 \pm 0.05$ & $0.08 \pm 0.02$ & $0.03 \pm 0.03$  & $0.59 \pm 0.03$ & $0.07 \pm 0.02$ & $0.01 \pm 0.01$ & $0.33 \pm 0.04$ & $0.04 \pm 0.03$ \\
{[}11,12{]}                                & $0.03 \pm 0.02$ & $0.62 \pm 0.07$ & $0.10 \pm 0.02$ & $-0.04 \pm 0.02$ & $0.81 \pm 0.08$ & $0.15 \pm 0.01$ & $0.05 \pm 0.03$ & $0.05 \pm 0.06$ & $0.03 \pm 0.03$ \\ \hline
\end{tabular}
}
\end{table*}

In general, we observe consistent patterns with those presented in Table \ref{tab:q_e_s0_model}. Notably, $a_{1}$ is approximately zero, independent of stellar mass, while $a_{2}$ declines with stellar mass. The $b$ parameters increase with stellar mass in both the quenched and early-type fractions. However, for the $F_{\rm S0,S}$ ratio, the $b$ parameter reaches a maximum value at intermediate stellar masses (purple and green curves in panel c). These trends indicate that massive systems have largely evolved before entering the cluster. Thus, we posit that most cluster-driven morphological transitions in the local universe occur in intermediate and low stellar mass galaxies.

By comparing the ratio of the rate of increase of $F_{\rm S0,S}$ to the early-type fraction (i.e., the ratio of the corresponding $a_{2}$ values), we gain insight into whether galaxies evolve towards elliptical morphologies or retain a remnant disk. Focusing on the two lower stellar mass bins, we find that $71 \pm 5 \%$ and $83 \pm 4 \%$ of galaxies undergoing morphological transition do so by transforming into S0s. While these fractions depend slightly on the exact T--type boundaries used to classify morphological types, the results highlight a significant distinction in the origins of cluster S0s -- environmentally-driven transformation of relatively low-mass infalling spirals -- and the origins of massive elliptical galaxies, which were likely present in the cores of clusters from earlier times.

The time-scales for star formation suppression and morphological transition can be compared using the $a_{2}$ slopes corresponding to the quenched and early-type fractions. Although the $a_{2}$ slopes are very similar in both cases, the initial $b$ parameter differs significantly. To estimate the time-scale for 50\% of the galaxies to be quenched or transformed into early-type, we use the equation $\tau_{50\%} = (0.5 - b)/a_{2}$, where $b$ and $a_{2}$ represent the results for quenching or morphological transformation in a given stellar mass bin. For the lowest stellar mass bins, it takes approximately, after the delay time, $2.4$ Gyr for star formation suppression and $1.2$ Gyr for morphological transition. This suggests a potential causal connection between these processes, with the removal of spiral arms being a critical step in the suppression of star formation.

\section{Conclusions}

In this paper, we have explored the impact of environmental factors on the suppression of star formation and the morphological transformation of galaxies infalling into clusters, with a particular emphasis on estimating the time-scales for these processes. Our primary focus has been on cluster galaxies and those entering these environments, while also including a comparison sample of field galaxies that evolve in isolation. Our results indicate that cluster galaxies, particularly those with lower stellar masses, are transitioning from the blue cloud to the green valley and red sequence due to a combination of environmental mechanisms. This environmental influence is also reflected in the morphology of these galaxies, as low-mass field galaxies exhibit higher T--Type values (later morphologies) compared to their cluster counterparts at the same position in the Star Formation Main Sequence diagram.

By examining the morphology of galaxies as a function of their deviation from the Star Formation Main Sequence (Fig.~\ref{fig:tinf_vs_deltaSFMS}), we reveal a strong correlation between T--Type and the perpendicular distance from the star formation main sequence ($\Delta \rm SFMS$). Our findings indicate that galaxies emerge from the green valley with early-type morphologies (T--Type < 0), regardless of their environment and stellar mass. This strongly suggests a close physical link between morphological transformation and the quenching of star formation.

To incorporate time into our analysis, we leverage the statistical relationship between a galaxy's position in the Projected Phase Space and the time since its infall, defined as the time elapsed since it first crossed $R_{200}$. Notably, the transition from the blue cloud to the red sequence corresponds to an increase in the average time since infall (Figs.~\ref{fig:PPS_distribution} and \ref{fig:tinf_vs_deltaSFMS}). Key findings from our time-scale analysis using the Projected Phase Space include:

\begin{itemize}

    \item By analyzing the distribution of blue cloud, green valley, and red sequence galaxies across the Projected Phase Space and as a function of infall time (Figs.~\ref{fig:pps_bc_gv_rs} and \ref{fig:tinf_bc_gv_rs}), we provide evidence supporting the slow-then-rapid quenching model \citep{2012MNRAS.424..232W, 2019ApJ...873...42R}. Notably, the variation in the fractions of galaxies at different evolutionary stages primarily occurs after a delay time of approximately 3.8 Gyr since infall. We claim that this delay corresponds to the time required for galaxies to reach an intracluster medium density sufficient for efficient ram pressure stripping of their gas \citep{2019ApJ...873...42R}. Similar results are obtained when considering morphological transformation, suggesting that the slow-then-rapid model also applies to morphological transition, reinforcing a possible connection between star formation suppression and morphological evolution;

    \item Our analysis shows an approximately constant fraction of green valley galaxies as a function of infall time, suggesting a balanced flow of galaxies into and out of this class as they fall into the cluster. Using a simple model to quantify the observed trends, we find that the most substantial variation occurs in low stellar mass galaxies, yielding a time-scale of approximately 0.5 Gyr for their transition through the green valley;

    \item By examining the fraction of quenched and early-type galaxies in the Projected Phase Space, our analysis reveals notable differences that become more pronounced with decreasing stellar mass (Figs.~\ref{fig:Quenched_Fraction_PPS}). In the lower stellar mass bin, we observe a higher early-type fraction at $0.3 < R_{\rm proj}/R_{200} < 1$ compared to the quenched fraction (35\% vs. 15\%), supporting the notion that morphological transition precedes full star formation quenching;

    \item Comparing the trends between the early-type fraction and the $F_{\rm S0,S} = N_{\rm S0}/(N_{\rm S0} + N_{\rm S})$ ratio across different masses and environments, we suggest that the increase in early-type galaxies in the two lower stellar mass bins is primarily due to a rise in the number of S0 galaxies relative to spirals. This highlights significant differences in morphological transitions between low and high mass galaxies. While massive ellipticals appear to be remnants of major merger events at higher redshifts, environmental effects such as ram pressure stripping on low-mass spiral galaxies transform them into S0s.

\end{itemize}

In summary, we demonstrate that the slow-then-rapid quenching model is applicable to both the suppression of star formation and the morphological transformation of galaxies infalling into clusters, possibly due to a shared causal connection between these processes. By quantifying time-scales using the relationship between location in the Projected Phase Space and time since infall, we provide a first, dynamically motivated estimate of the time galaxies spend in the green valley due to environmental effects. Our analysis also reveals that the fractions of quenched and early-type galaxies remain nearly unchanged until a delay time of $t_{\rm delay} \sim 3.8$ Gyr. We estimate that it takes approximately 2.4 and 1.2 Gyr, respectively, after the delay time for the fractions of quenched and early-type galaxies in clusters to surpass 50\% in low-mass galaxies. These time-scales reinforce the notion that morphological transition precedes complete star formation quenching and suggest that the physical mechanisms driving these changes may be interconnected.

\section*{Acknowledgements}

VMS and RRdC acknowledge the support from FAPESP through the grants 2020/16243-3 and 2020/15245-2. MRM, AAS and DJC acknowledge financial support from the UK Science and Technology Facilities Council (STFC; grant ref: ST/T000171/1, and through a PhD studentship). IF acknowledges support from the Spanish Ministry of Science, Innovation and Universities (MCIU), through grant PID2019-104788GB-I00. VMS and RRdC thanks M. Oxland and L. C. Parker for constructive discussions on the subject.

Funding for the Sloan Digital Sky Survey V has been provided by the Alfred P. Sloan Foundation, the Heising-Simons Foundation, the National Science Foundation, and the Participating Institutions. SDSS acknowledges support and resources from the Center for High-Performance Computing at the University of Utah. SDSS telescopes are located at Apache Point Observatory, funded by the Astrophysical Research Consortium and operated by New Mexico State University, and at Las Campanas Observatory, operated by the Carnegie Institution for Science. The SDSS web site is www.sdss.org.

SDSS is managed by the Astrophysical Research Consortium for the Participating Institutions of the SDSS Collaboration, including Caltech, the Carnegie Institution for Science, Chilean National Time Allocation Committee (CNTAC) ratified researchers, The Flatiron Institute, the Gotham Participation Group, Harvard University, Heidelberg University, The Johns Hopkins University, L’Ecole polytechnique fédérale de Lausanne (EPFL), Leibniz-Institut für Astrophysik Potsdam (AIP), Max-Planck-Institut für Astronomie (MPIA Heidelberg), Max-Planck-Institut für Extraterrestrische Physik (MPE), Nanjing University, National Astronomical Observatories of China (NAOC), New Mexico State University, The Ohio State University, Pennsylvania State University, Smithsonian Astrophysical Observatory, Space Telescope Science Institute (STScI), the Stellar Astrophysics Participation Group, Universidad Nacional Autónoma de México, University of Arizona, University of Colorado Boulder, University of Illinois at Urbana-Champaign, University of Toronto, University of Utah, University of Virginia, Yale University, and Yunnan University.

\section*{Data Availability}

The data underlying this paper were accessed from SDSS DR16 database
(https://skyserver.sdss.org/dr16/en/home). The data underlying this article will be shared on request to the corresponding author.



\bibliographystyle{mnras}
\bibliography{example} 

\begin{thebibliography}{}
\makeatletter
\relax
\def\mn@urlcharsother{\let\do\@makeother \do\$\do\&\do\#\do\^\do\_\do\%\do\~}
\def\mn@doi{\begingroup\mn@urlcharsother \@ifnextchar [ {\mn@doi@} {\mn@doi@[]}}
\def\mn@doi@[#1]#2{\def\@tempa{#1}\ifx\@tempa\@empty \href {http://dx.doi.org/#2} {doi:#2}\else \href {http://dx.doi.org/#2} {#1}\fi \endgroup}
\def\mn@eprint#1#2{\mn@eprint@#1:#2::\@nil}
\def\mn@eprint@arXiv#1{\href {http://arxiv.org/abs/#1} {{\tt arXiv:#1}}}
\def\mn@eprint@dblp#1{\href {http://dblp.uni-trier.de/rec/bibtex/#1.xml} {dblp:#1}}
\def\mn@eprint@#1:#2:#3:#4\@nil{\def\@tempa {#1}\def\@tempb {#2}\def\@tempc {#3}\ifx \@tempc \@empty \let \@tempc \@tempb \let \@tempb \@tempa \fi \ifx \@tempb \@empty \def\@tempb {arXiv}\fi \@ifundefined {mn@eprint@\@tempb}{\@tempb:\@tempc}{\expandafter \expandafter \csname mn@eprint@\@tempb\endcsname \expandafter{\@tempc}}}

\bibitem[\protect\citeauthoryear{{Abadi}, {Moore}  \& {Bower}}{{Abadi} et~al.}{1999}]{1999MNRAS.308..947A}
{Abadi} M.~G.,  {Moore} B.,   {Bower} R.~G.,  1999, \mn@doi [\mnras] {10.1046/j.1365-8711.1999.02715.x}, \href {https://ui.adsabs.harvard.edu/abs/1999MNRAS.308..947A} {308, 947}

\bibitem[\protect\citeauthoryear{{Ahumada} et~al.,}{{Ahumada} et~al.}{2020}]{2020ApJS..249....3A}
{Ahumada} R.,  et~al., 2020, \mn@doi [\apjs] {10.3847/1538-4365/ab929e}, \href {https://ui.adsabs.harvard.edu/abs/2020ApJS..249....3A} {249, 3}

\bibitem[\protect\citeauthoryear{{Arnouts} et~al.,}{{Arnouts} et~al.}{2007}]{2007A&A...476..137A}
{Arnouts} S.,  et~al., 2007, \mn@doi [\aap] {10.1051/0004-6361:20077632}, \href {https://ui.adsabs.harvard.edu/abs/2007A&A...476..137A} {476, 137}

\bibitem[\protect\citeauthoryear{{Balogh}, {Navarro}  \& {Morris}}{{Balogh} et~al.}{2000}]{2000ApJ...540..113B}
{Balogh} M.~L.,  {Navarro} J.~F.,   {Morris} S.~L.,  2000, \mn@doi [\apj] {10.1086/309323}, \href {https://ui.adsabs.harvard.edu/abs/2000ApJ...540..113B} {540, 113}

\bibitem[\protect\citeauthoryear{{Blanton} et~al.,}{{Blanton} et~al.}{2005}]{2005AJ....129.2562B}
{Blanton} M.~R.,  et~al., 2005, \mn@doi [\aj] {10.1086/429803}, \href {https://ui.adsabs.harvard.edu/abs/2005AJ....129.2562B} {129, 2562}

\bibitem[\protect\citeauthoryear{{Bluck} et~al.,}{{Bluck} et~al.}{2020}]{2020MNRAS.499..230B}
{Bluck} A. F.~L.,  et~al., 2020, \mn@doi [\mnras] {10.1093/mnras/staa2806}, \href {https://ui.adsabs.harvard.edu/abs/2020MNRAS.499..230B} {499, 230}

\bibitem[\protect\citeauthoryear{{Bongiorno} et~al.,}{{Bongiorno} et~al.}{2016}]{Bongiorno}
{Bongiorno} A.,  et~al., 2016, \mn@doi [\aap] {10.1051/0004-6361/201527436}, \href {https://ui.adsabs.harvard.edu/abs/2016A&A...588A..78B} {588, A78}

\bibitem[\protect\citeauthoryear{{Brinchmann}, {Charlot}, {White}, {Tremonti}, {Kauffmann}, {Heckman}  \& {Brinkmann}}{{Brinchmann} et~al.}{2004}]{2004MNRAS.351.1151B}
{Brinchmann} J.,  {Charlot} S.,  {White} S.~D.~M.,  {Tremonti} C.,  {Kauffmann} G.,  {Heckman} T.,   {Brinkmann} J.,  2004, \mn@doi [\mnras] {10.1111/j.1365-2966.2004.07881.x}, \href {https://ui.adsabs.harvard.edu/abs/2004MNRAS.351.1151B} {351, 1151}

\bibitem[\protect\citeauthoryear{{Capelato}, {Gerbal}, {Salvador-Sole}, {Mathez}, {Mazure}  \& {Sol}}{{Capelato} et~al.}{1980}]{1980ApJ...241..521C}
{Capelato} H.~V.,  {Gerbal} D.,  {Salvador-Sole} E.,  {Mathez} G.,  {Mazure} A.,   {Sol} H.,  1980, \mn@doi [\apj] {10.1086/158366}, \href {https://ui.adsabs.harvard.edu/abs/1980ApJ...241..521C} {241, 521}

\bibitem[\protect\citeauthoryear{{Carlberg} et~al.,}{{Carlberg} et~al.}{1997}]{1997ApJ...476L...7C}
{Carlberg} R.~G.,  et~al., 1997, \mn@doi [\apjl] {10.1086/310497}, \href {https://ui.adsabs.harvard.edu/abs/1997ApJ...476L...7C} {476, L7}

\bibitem[\protect\citeauthoryear{{Choi} \& {Yi}}{{Choi} \& {Yi}}{2017}]{2017ApJ...837...68C}
{Choi} H.,  {Yi} S.~K.,  2017, \mn@doi [\apj] {10.3847/1538-4357/aa5e4b}, \href {https://ui.adsabs.harvard.edu/abs/2017ApJ...837...68C} {837, 68}

\bibitem[\protect\citeauthoryear{{Cimatti}, {Fraternali}  \& {Nipoti}}{{Cimatti} et~al.}{2019}]{2019igfe.book.....C}
{Cimatti} A.,  {Fraternali} F.,   {Nipoti} C.,  2019, {Introduction to Galaxy Formation and Evolution: From Primordial Gas to Present-Day Galaxies}.
Cambridge University Press

\bibitem[\protect\citeauthoryear{{Cooper} et~al.,}{{Cooper} et~al.}{2022}]{2022MNRAS.509.5382C}
{Cooper} J.~R.,  et~al., 2022, \mn@doi [\mnras] {10.1093/mnras/stab3184}, \href {https://ui.adsabs.harvard.edu/abs/2022MNRAS.509.5382C} {509, 5382}

\bibitem[\protect\citeauthoryear{{Cox}, {Jonsson}, {Somerville}, {Primack}  \& {Dekel}}{{Cox} et~al.}{2008}]{2008MNRAS.384..386C}
{Cox} T.~J.,  {Jonsson} P.,  {Somerville} R.~S.,  {Primack} J.~R.,   {Dekel} A.,  2008, \mn@doi [\mnras] {10.1111/j.1365-2966.2007.12730.x}, \href {https://ui.adsabs.harvard.edu/abs/2008MNRAS.384..386C} {384, 386}

\bibitem[\protect\citeauthoryear{{Dalla Vecchia} \& {Schaye}}{{Dalla Vecchia} \& {Schaye}}{2008}]{2008MNRAS.387.1431D}
{Dalla Vecchia} C.,  {Schaye} J.,  2008, \mn@doi [\mnras] {10.1111/j.1365-2966.2008.13322.x}, \href {https://ui.adsabs.harvard.edu/abs/2008MNRAS.387.1431D} {387, 1431}

\bibitem[\protect\citeauthoryear{{Davidzon} et~al.,}{{Davidzon} et~al.}{2016}]{2016A&A...586A..23D}
{Davidzon} I.,  et~al., 2016, \mn@doi [\aap] {10.1051/0004-6361/201527129}, \href {https://ui.adsabs.harvard.edu/abs/2016A&A...586A..23D} {586, A23}

\bibitem[\protect\citeauthoryear{{Dom{\'\i}nguez S{\'a}nchez}, {Huertas-Company}, {Bernardi}, {Tuccillo}  \& {Fischer}}{{Dom{\'\i}nguez S{\'a}nchez} et~al.}{2018}]{2018MNRAS.476.3661D}
{Dom{\'\i}nguez S{\'a}nchez} H.,  {Huertas-Company} M.,  {Bernardi} M.,  {Tuccillo} D.,   {Fischer} J.~L.,  2018, \mn@doi [\mnras] {10.1093/mnras/sty338}, \href {https://ui.adsabs.harvard.edu/abs/2018MNRAS.476.3661D} {476, 3661}

\bibitem[\protect\citeauthoryear{{Dressler}}{{Dressler}}{1980}]{Dressler}
{Dressler} A.,  1980, \mn@doi [\apj] {10.1086/157753}, \href {https://ui.adsabs.harvard.edu/abs/1980ApJ...236..351D} {236, 351}

\bibitem[\protect\citeauthoryear{{Dressler}}{{Dressler}}{1996}]{1996hst..prop.6480D}
{Dressler} A.,  1996, {The Butcher-Oemler Effect: where have all the Spirals Gone?}, HST Proposal ID 6480. Cycle 6

\bibitem[\protect\citeauthoryear{{Dressler} \& {Smail}}{{Dressler} \& {Smail}}{1997}]{1997hsth.conf..185D}
{Dressler} A.,  {Smail} I.,  1997, in {Tanvir} N.~R.,  {Aragon-Salamanca} A.,   {Wall} J.~V.,  eds, The Hubble Space Telescope and the High Redshift Universe. p.~185 (\mn@eprint {arXiv} {astro-ph/9611004}), \mn@doi{10.48550/arXiv.astro-ph/9611004}

\bibitem[\protect\citeauthoryear{{Ellingson}, {Lin}, {Yee}  \& {Carlberg}}{{Ellingson} et~al.}{2001}]{2001ApJ...547..609E}
{Ellingson} E.,  {Lin} H.,  {Yee} H.~K.~C.,   {Carlberg} R.~G.,  2001, \mn@doi [\apj] {10.1086/318423}, \href {https://ui.adsabs.harvard.edu/abs/2001ApJ...547..609E} {547, 609}

\bibitem[\protect\citeauthoryear{{Ferreras}, {B{\"o}hm}, {Umetsu}, {Sampaio}  \& {de Carvalho}}{{Ferreras} et~al.}{2023}]{2023MNRAS.519.4884F}
{Ferreras} I.,  {B{\"o}hm} A.,  {Umetsu} K.,  {Sampaio} V.,   {de Carvalho} R.~R.,  2023, \mn@doi [\mnras] {10.1093/mnras/stad001}, \href {https://ui.adsabs.harvard.edu/abs/2023MNRAS.519.4884F} {519, 4884}

\bibitem[\protect\citeauthoryear{{Fujita}}{{Fujita}}{2004}]{2004PASJ...56...29F}
{Fujita} Y.,  2004, \mn@doi [\pasj] {10.1093/pasj/56.1.29}, \href {https://ui.adsabs.harvard.edu/abs/2004PASJ...56...29F} {56, 29}

\bibitem[\protect\citeauthoryear{{Gunn} \& {Gott}}{{Gunn} \& {Gott}}{1972}]{1972ApJ...176....1G}
{Gunn} J.~E.,  {Gott} J.~Richard I.,  1972, \mn@doi [\apj] {10.1086/151605}, \href {https://ui.adsabs.harvard.edu/abs/1972ApJ...176....1G} {176, 1}

\bibitem[\protect\citeauthoryear{{Haggar}, {Gray}, {Pearce}, {Knebe}, {Cui}, {Mostoghiu}  \& {Yepes}}{{Haggar} et~al.}{2020}]{2020MNRAS.492.6074H}
{Haggar} R.,  {Gray} M.~E.,  {Pearce} F.~R.,  {Knebe} A.,  {Cui} W.,  {Mostoghiu} R.,   {Yepes} G.,  2020, \mn@doi [\mnras] {10.1093/mnras/staa273}, \href {https://ui.adsabs.harvard.edu/abs/2020MNRAS.492.6074H} {492, 6074}

\bibitem[\protect\citeauthoryear{{Hart} et~al.,}{{Hart} et~al.}{2016}]{2016MNRAS.461.3663H}
{Hart} R.~E.,  et~al., 2016, \mn@doi [\mnras] {10.1093/mnras/stw1588}, \href {https://ui.adsabs.harvard.edu/abs/2016MNRAS.461.3663H} {461, 3663}

\bibitem[\protect\citeauthoryear{{Johnston}, {Sigurdsson}  \& {Hernquist}}{{Johnston} et~al.}{1999}]{1999MNRAS.302..771J}
{Johnston} K.~V.,  {Sigurdsson} S.,   {Hernquist} L.,  1999, \mn@doi [\mnras] {10.1046/j.1365-8711.1999.02200.x}, \href {https://ui.adsabs.harvard.edu/abs/1999MNRAS.302..771J} {302, 771}

\bibitem[\protect\citeauthoryear{{Kauffmann} et~al.,}{{Kauffmann} et~al.}{2003}]{2003MNRAS.341...33K}
{Kauffmann} G.,  et~al., 2003, \mn@doi [\mnras] {10.1046/j.1365-8711.2003.06291.x}, \href {https://ui.adsabs.harvard.edu/abs/2003MNRAS.341...33K} {341, 33}

\bibitem[\protect\citeauthoryear{{Kelkar}, {Gray}, {Arag{\'o}n-Salamanca}, {Rudnick}, {Jaff{\'e}}, {Jablonka}, {Moustakas}  \& {Milvang-Jensen}}{{Kelkar} et~al.}{2019}]{2019MNRAS.486..868K}
{Kelkar} K.,  {Gray} M.~E.,  {Arag{\'o}n-Salamanca} A.,  {Rudnick} G.,  {Jaff{\'e}} Y.~L.,  {Jablonka} P.,  {Moustakas} J.,   {Milvang-Jensen} B.,  2019, \mn@doi [\mnras] {10.1093/mnras/stz905}, \href {https://ui.adsabs.harvard.edu/abs/2019MNRAS.486..868K} {486, 868}

\bibitem[\protect\citeauthoryear{{Kelvin} et~al.,}{{Kelvin} et~al.}{2018}]{2018MNRAS.477.4116K}
{Kelvin} L.~S.,  et~al., 2018, \mn@doi [\mnras] {10.1093/mnras/sty933}, \href {https://ui.adsabs.harvard.edu/abs/2018MNRAS.477.4116K} {477, 4116}

\bibitem[\protect\citeauthoryear{{La Barbera}, {de Carvalho}, {de la Rosa}, {Sorrentino}, {Gal}  \& {Kohl-Moreira}}{{La Barbera} et~al.}{2009}]{2009AJ....137.3942L}
{La Barbera} F.,  {de Carvalho} R.~R.,  {de la Rosa} I.~G.,  {Sorrentino} G.,  {Gal} R.~R.,   {Kohl-Moreira} J.~L.,  2009, \mn@doi [\aj] {10.1088/0004-6256/137/4/3942}, \href {https://ui.adsabs.harvard.edu/abs/2009AJ....137.3942L} {137, 3942}

\bibitem[\protect\citeauthoryear{{Labb{\'e}} et~al.,}{{Labb{\'e}} et~al.}{2005}]{2005ApJ...624L..81L}
{Labb{\'e}} I.,  et~al., 2005, \mn@doi [\apjl] {10.1086/430700}, \href {https://ui.adsabs.harvard.edu/abs/2005ApJ...624L..81L} {624, L81}

\bibitem[\protect\citeauthoryear{{Larson}, {Tinsley}  \& {Caldwell}}{{Larson} et~al.}{1980}]{1980ApJ...237..692L}
{Larson} R.~B.,  {Tinsley} B.~M.,   {Caldwell} C.~N.,  1980, \mn@doi [\apj] {10.1086/157917}, \href {https://ui.adsabs.harvard.edu/abs/1980ApJ...237..692L} {237, 692}

\bibitem[\protect\citeauthoryear{{Lopes}, {de Carvalho}, {Kohl-Moreira}  \& {Jones}}{{Lopes} et~al.}{2009}]{2009MNRAS.399.2201L}
{Lopes} P.~A.~A.,  {de Carvalho} R.~R.,  {Kohl-Moreira} J.~L.,   {Jones} C.,  2009, \mn@doi [\mnras] {10.1111/j.1365-2966.2009.15425.x}, \href {https://ui.adsabs.harvard.edu/abs/2009MNRAS.399.2201L} {399, 2201}

\bibitem[\protect\citeauthoryear{{Madau} \& {Dickinson}}{{Madau} \& {Dickinson}}{2014}]{2014ARA&A..52..415M}
{Madau} P.,  {Dickinson} M.,  2014, \mn@doi [\araa] {10.1146/annurev-astro-081811-125615}, \href {https://ui.adsabs.harvard.edu/abs/2014ARA&A..52..415M} {52, 415}

\bibitem[\protect\citeauthoryear{{Mahajan}}{{Mahajan}}{2013}]{2013MNRAS.431L.117M}
{Mahajan} S.,  2013, \mn@doi [\mnras] {10.1093/mnrasl/slt021}, \href {https://ui.adsabs.harvard.edu/abs/2013MNRAS.431L.117M} {431, L117}

\bibitem[\protect\citeauthoryear{{Mahajan}, {Mamon}  \& {Raychaudhury}}{{Mahajan} et~al.}{2011}]{2011MNRAS.416.2882M}
{Mahajan} S.,  {Mamon} G.~A.,   {Raychaudhury} S.,  2011, \mn@doi [\mnras] {10.1111/j.1365-2966.2011.19236.x}, \href {https://ui.adsabs.harvard.edu/abs/2011MNRAS.416.2882M} {416, 2882}

\bibitem[\protect\citeauthoryear{{Mancini} et~al.,}{{Mancini} et~al.}{2019}]{2019MNRAS.489.1265M}
{Mancini} C.,  et~al., 2019, \mn@doi [\mnras] {10.1093/mnras/stz2130}, \href {https://ui.adsabs.harvard.edu/abs/2019MNRAS.489.1265M} {489, 1265}

\bibitem[\protect\citeauthoryear{{Mapelli}}{{Mapelli}}{2015}]{2015Galax...3..192M}
{Mapelli} M.,  2015, \mn@doi [Galaxies] {10.3390/galaxies3040192}, \href {https://ui.adsabs.harvard.edu/abs/2015Galax...3..192M} {3, 192}

\bibitem[\protect\citeauthoryear{{Martig}, {Bournaud}, {Teyssier}  \& {Dekel}}{{Martig} et~al.}{2009}]{2009ApJ...707..250M}
{Martig} M.,  {Bournaud} F.,  {Teyssier} R.,   {Dekel} A.,  2009, \mn@doi [\apj] {10.1088/0004-637X/707/1/250}, \href {https://ui.adsabs.harvard.edu/abs/2009ApJ...707..250M} {707, 250}

\bibitem[\protect\citeauthoryear{{Masters} et~al.,}{{Masters} et~al.}{2010a}]{2010MNRAS.404..792M}
{Masters} K.~L.,  et~al., 2010a, \mn@doi [\mnras] {10.1111/j.1365-2966.2010.16335.x}, \href {https://ui.adsabs.harvard.edu/abs/2010MNRAS.404..792M} {404, 792}

\bibitem[\protect\citeauthoryear{{Masters} et~al.,}{{Masters} et~al.}{2010b}]{2010MNRAS.405..783M}
{Masters} K.~L.,  et~al., 2010b, \mn@doi [\mnras] {10.1111/j.1365-2966.2010.16503.x}, \href {https://ui.adsabs.harvard.edu/abs/2010MNRAS.405..783M} {405, 783}

\bibitem[\protect\citeauthoryear{{McNab} et~al.,}{{McNab} et~al.}{2021}]{2021MNRAS.508..157M}
{McNab} K.,  et~al., 2021, \mn@doi [\mnras] {10.1093/mnras/stab2558}, \href {https://ui.adsabs.harvard.edu/abs/2021MNRAS.508..157M} {508, 157}

\bibitem[\protect\citeauthoryear{{Meert}, {Vikram}  \& {Bernardi}}{{Meert} et~al.}{2015}]{2015MNRAS.446.3943M}
{Meert} A.,  {Vikram} V.,   {Bernardi} M.,  2015, \mn@doi [\mnras] {10.1093/mnras/stu2333}, \href {https://ui.adsabs.harvard.edu/abs/2015MNRAS.446.3943M} {446, 3943}

\bibitem[\protect\citeauthoryear{{Moutard} et~al.,}{{Moutard} et~al.}{2016}]{2016A&A...590A.102M}
{Moutard} T.,  et~al., 2016, \mn@doi [\aap] {10.1051/0004-6361/201527945}, \href {https://ui.adsabs.harvard.edu/abs/2016A&A...590A.102M} {590, A102}

\bibitem[\protect\citeauthoryear{{Oman} \& {Hudson}}{{Oman} \& {Hudson}}{2016}]{2016MNRAS.463.3083O}
{Oman} K.~A.,  {Hudson} M.~J.,  2016, \mn@doi [\mnras] {10.1093/mnras/stw2195}, \href {https://ui.adsabs.harvard.edu/abs/2016MNRAS.463.3083O} {463, 3083}

\bibitem[\protect\citeauthoryear{{Oman}, {Hudson}  \& {Behroozi}}{{Oman} et~al.}{2013}]{2013MNRAS.431.2307O}
{Oman} K.~A.,  {Hudson} M.~J.,   {Behroozi} P.~S.,  2013, \mn@doi [\mnras] {10.1093/mnras/stt328}, \href {https://ui.adsabs.harvard.edu/abs/2013MNRAS.431.2307O} {431, 2307}

\bibitem[\protect\citeauthoryear{{Pasquali}, {Smith}, {Gallazzi}, {De Lucia}, {Zibetti}, {Hirschmann}  \& {Yi}}{{Pasquali} et~al.}{2019}]{2019MNRAS.484.1702P}
{Pasquali} A.,  {Smith} R.,  {Gallazzi} A.,  {De Lucia} G.,  {Zibetti} S.,  {Hirschmann} M.,   {Yi} S.~K.,  2019, \mn@doi [\mnras] {10.1093/mnras/sty3530}, \href {https://ui.adsabs.harvard.edu/abs/2019MNRAS.484.1702P} {484, 1702}

\bibitem[\protect\citeauthoryear{{Peng} et~al.,}{{Peng} et~al.}{2010}]{2010ApJ...721..193P}
{Peng} Y.-j.,  et~al., 2010, \mn@doi [\apj] {10.1088/0004-637X/721/1/193}, \href {https://ui.adsabs.harvard.edu/abs/2010ApJ...721..193P} {721, 193}

\bibitem[\protect\citeauthoryear{{Read}, {Wilkinson}, {Evans}, {Gilmore}  \& {Kleyna}}{{Read} et~al.}{2006}]{2006MNRAS.366..429R}
{Read} J.~I.,  {Wilkinson} M.~I.,  {Evans} N.~W.,  {Gilmore} G.,   {Kleyna} J.~T.,  2006, \mn@doi [\mnras] {10.1111/j.1365-2966.2005.09861.x}, \href {https://ui.adsabs.harvard.edu/abs/2006MNRAS.366..429R} {366, 429}

\bibitem[\protect\citeauthoryear{{Rhee}, {Smith}, {Choi}, {Yi}, {Jaff{\'e}}, {Candlish}  \& {S{\'a}nchez-J{\'a}nssen}}{{Rhee} et~al.}{2017}]{2017ApJ...843..128R}
{Rhee} J.,  {Smith} R.,  {Choi} H.,  {Yi} S.~K.,  {Jaff{\'e}} Y.,  {Candlish} G.,   {S{\'a}nchez-J{\'a}nssen} R.,  2017, \mn@doi [\apj] {10.3847/1538-4357/aa6d6c}, \href {https://ui.adsabs.harvard.edu/abs/2017ApJ...843..128R} {843, 128}

\bibitem[\protect\citeauthoryear{{Rhee}, {Smith}, {Choi}, {Contini}, {Jung}, {Han}  \& {Yi}}{{Rhee} et~al.}{2020}]{2020ApJS..247...45R}
{Rhee} J.,  {Smith} R.,  {Choi} H.,  {Contini} E.,  {Jung} S.~L.,  {Han} S.,   {Yi} S.~K.,  2020, \mn@doi [\apjs] {10.3847/1538-4365/ab7377}, \href {https://ui.adsabs.harvard.edu/abs/2020ApJS..247...45R} {247, 45}

\bibitem[\protect\citeauthoryear{{Ribeiro}, {de Carvalho}, {Trevisan}, {Capelato}, {La Barbera}, {Lopes}  \& {Schilling}}{{Ribeiro} et~al.}{2013}]{2013MNRAS.434..784R}
{Ribeiro} A.~L.~B.,  {de Carvalho} R.~R.,  {Trevisan} M.,  {Capelato} H.~V.,  {La Barbera} F.,  {Lopes} P.~A.~A.,   {Schilling} A.~C.,  2013, \mn@doi [\mnras] {10.1093/mnras/stt1071}, \href {https://ui.adsabs.harvard.edu/abs/2013MNRAS.434..784R} {434, 784}

\bibitem[\protect\citeauthoryear{{Roberts} \& {Parker}}{{Roberts} \& {Parker}}{2017}]{2017MNRAS.467.3268R}
{Roberts} I.~D.,  {Parker} L.~C.,  2017, \mn@doi [\mnras] {10.1093/mnras/stx317}, \href {https://ui.adsabs.harvard.edu/abs/2017MNRAS.467.3268R} {467, 3268}

\bibitem[\protect\citeauthoryear{{Roberts} \& {Parker}}{{Roberts} \& {Parker}}{2019}]{2019MNRAS.490..773R}
{Roberts} I.~D.,  {Parker} L.~C.,  2019, \mn@doi [\mnras] {10.1093/mnras/stz2666}, \href {https://ui.adsabs.harvard.edu/abs/2019MNRAS.490..773R} {490, 773}

\bibitem[\protect\citeauthoryear{{Roberts} \& {Parker}}{{Roberts} \& {Parker}}{2020}]{2020MNRAS.495..554R}
{Roberts} I.~D.,  {Parker} L.~C.,  2020, \mn@doi [\mnras] {10.1093/mnras/staa1213}, \href {https://ui.adsabs.harvard.edu/abs/2020MNRAS.495..554R} {495, 554}

\bibitem[\protect\citeauthoryear{{Roberts}, {Parker}, {Brown}, {Joshi}, {Hlavacek-Larrondo}  \& {Wadsley}}{{Roberts} et~al.}{2019}]{2019ApJ...873...42R}
{Roberts} I.~D.,  {Parker} L.~C.,  {Brown} T.,  {Joshi} G.~D.,  {Hlavacek-Larrondo} J.,   {Wadsley} J.,  2019, \mn@doi [\apj] {10.3847/1538-4357/ab04f7}, \href {https://ui.adsabs.harvard.edu/abs/2019ApJ...873...42R} {873, 42}

\bibitem[\protect\citeauthoryear{{Salim} et~al.,}{{Salim} et~al.}{2007}]{2007ApJS..173..267S}
{Salim} S.,  et~al., 2007, \mn@doi [\apjs] {10.1086/519218}, \href {https://ui.adsabs.harvard.edu/abs/2007ApJS..173..267S} {173, 267}

\bibitem[\protect\citeauthoryear{{Sampaio}, {de Carvalho}, {Ferreras}, {Lagan{\'a}}, {Ribeiro}  \& {Rembold}}{{Sampaio} et~al.}{2021}]{2021MNRAS.503.3065S}
{Sampaio} V.~M.,  {de Carvalho} R.~R.,  {Ferreras} I.,  {Lagan{\'a}} T.~F.,  {Ribeiro} A.~L.~B.,   {Rembold} S.~B.,  2021, \mn@doi [\mnras] {10.1093/mnras/stab673}, \href {https://ui.adsabs.harvard.edu/abs/2021MNRAS.503.3065S} {503, 3065}

\bibitem[\protect\citeauthoryear{{Sampaio}, {de Carvalho}, {Ferreras}, {Arag{\'o}n-Salamanca}  \& {Parker}}{{Sampaio} et~al.}{2022}]{2022MNRAS.509..567S}
{Sampaio} V.~M.,  {de Carvalho} R.~R.,  {Ferreras} I.,  {Arag{\'o}n-Salamanca} A.,   {Parker} L.~C.,  2022, \mn@doi [\mnras] {10.1093/mnras/stab3018}, \href {https://ui.adsabs.harvard.edu/abs/2022MNRAS.509..567S} {509, 567}

\bibitem[\protect\citeauthoryear{{Sampaio}, {Arag{\'o}n-Salamanca}, {Merrifield}, {de Carvalho}, {Zhou}  \& {Ferreras}}{{Sampaio} et~al.}{2023}]{2023MNRAS.524.5327S}
{Sampaio} V.~M.,  {Arag{\'o}n-Salamanca} A.,  {Merrifield} M.~R.,  {de Carvalho} R.~R.,  {Zhou} S.,   {Ferreras} I.,  2023, \mn@doi [\mnras] {10.1093/mnras/stad2211}, \href {https://ui.adsabs.harvard.edu/abs/2023MNRAS.524.5327S} {524, 5327}

\bibitem[\protect\citeauthoryear{{Samui}, {Subramanian}  \& {Srianand}}{{Samui} et~al.}{2018}]{2018MNRAS.476.1680S}
{Samui} S.,  {Subramanian} K.,   {Srianand} R.,  2018, \mn@doi [\mnras] {10.1093/mnras/sty287}, \href {https://ui.adsabs.harvard.edu/abs/2018MNRAS.476.1680S} {476, 1680}

\bibitem[\protect\citeauthoryear{{Sarron}, {Adami}, {Durret}  \& {Laigle}}{{Sarron} et~al.}{2019}]{2019A&A...632A..49S}
{Sarron} F.,  {Adami} C.,  {Durret} F.,   {Laigle} C.,  2019, \mn@doi [\aap] {10.1051/0004-6361/201935394}, \href {https://ui.adsabs.harvard.edu/abs/2019A&A...632A..49S} {632, A49}

\bibitem[\protect\citeauthoryear{{Schawinski} et~al.,}{{Schawinski} et~al.}{2009}]{2009MNRAS.396..818S}
{Schawinski} K.,  et~al., 2009, \mn@doi [\mnras] {10.1111/j.1365-2966.2009.14793.x}, \href {https://ui.adsabs.harvard.edu/abs/2009MNRAS.396..818S} {396, 818}

\bibitem[\protect\citeauthoryear{{Schawinski} et~al.,}{{Schawinski} et~al.}{2014}]{2014MNRAS.440..889S}
{Schawinski} K.,  et~al., 2014, \mn@doi [\mnras] {10.1093/mnras/stu327}, \href {https://ui.adsabs.harvard.edu/abs/2014MNRAS.440..889S} {440, 889}

\bibitem[\protect\citeauthoryear{{Spinoso}, {Bonoli}, {Dotti}, {Mayer}, {Madau}  \& {Bellovary}}{{Spinoso} et~al.}{2017}]{2017MNRAS.465.3729S}
{Spinoso} D.,  {Bonoli} S.,  {Dotti} M.,  {Mayer} L.,  {Madau} P.,   {Bellovary} J.,  2017, \mn@doi [\mnras] {10.1093/mnras/stw2934}, \href {https://ui.adsabs.harvard.edu/abs/2017MNRAS.465.3729S} {465, 3729}

\bibitem[\protect\citeauthoryear{{Springel} \& {Hernquist}}{{Springel} \& {Hernquist}}{2005}]{2005ApJ...622L...9S}
{Springel} V.,  {Hernquist} L.,  2005, \mn@doi [\apjl] {10.1086/429486}, \href {https://ui.adsabs.harvard.edu/abs/2005ApJ...622L...9S} {622, L9}

\bibitem[\protect\citeauthoryear{{Tempel}, {Stoica}, {Mart{\'\i}nez}, {Liivam{\"a}gi}, {Castellan}  \& {Saar}}{{Tempel} et~al.}{2014}]{2014MNRAS.438.3465T}
{Tempel} E.,  {Stoica} R.~S.,  {Mart{\'\i}nez} V.~J.,  {Liivam{\"a}gi} L.~J.,  {Castellan} G.,   {Saar} E.,  2014, \mn@doi [\mnras] {10.1093/mnras/stt2454}, \href {https://ui.adsabs.harvard.edu/abs/2014MNRAS.438.3465T} {438, 3465}

\bibitem[\protect\citeauthoryear{{Teyssier}, {Chapon}  \& {Bournaud}}{{Teyssier} et~al.}{2010}]{2010ApJ...720L.149T}
{Teyssier} R.,  {Chapon} D.,   {Bournaud} F.,  2010, \mn@doi [\apjl] {10.1088/2041-8205/720/2/L149}, \href {https://ui.adsabs.harvard.edu/abs/2010ApJ...720L.149T} {720, L149}

\bibitem[\protect\citeauthoryear{{Thilker} et~al.,}{{Thilker} et~al.}{2010}]{2010ApJ...714L.171T}
{Thilker} D.~A.,  et~al., 2010, \mn@doi [\apjl] {10.1088/2041-8205/714/1/L171}, \href {https://ui.adsabs.harvard.edu/abs/2010ApJ...714L.171T} {714, L171}

\bibitem[\protect\citeauthoryear{{Tomczak} et~al.,}{{Tomczak} et~al.}{2016}]{2016ApJ...817..118T}
{Tomczak} A.~R.,  et~al., 2016, \mn@doi [\apj] {10.3847/0004-637X/817/2/118}, \href {https://ui.adsabs.harvard.edu/abs/2016ApJ...817..118T} {817, 118}

\bibitem[\protect\citeauthoryear{{Trussler}, {Maiolino}, {Maraston}, {Peng}, {Thomas}, {Goddard}  \& {Lian}}{{Trussler} et~al.}{2020}]{2020MNRAS.491.5406T}
{Trussler} J.,  {Maiolino} R.,  {Maraston} C.,  {Peng} Y.,  {Thomas} D.,  {Goddard} D.,   {Lian} J.,  2020, \mn@doi [\mnras] {10.1093/mnras/stz3286}, \href {https://ui.adsabs.harvard.edu/abs/2020MNRAS.491.5406T} {491, 5406}

\bibitem[\protect\citeauthoryear{{Wetzel}, {Tinker}  \& {Conroy}}{{Wetzel} et~al.}{2012}]{2012MNRAS.424..232W}
{Wetzel} A.~R.,  {Tinker} J.~L.,   {Conroy} C.,  2012, \mn@doi [\mnras] {10.1111/j.1365-2966.2012.21188.x}, \href {https://ui.adsabs.harvard.edu/abs/2012MNRAS.424..232W} {424, 232}

\bibitem[\protect\citeauthoryear{{Wetzel}, {Tinker}, {Conroy}  \& {van den Bosch}}{{Wetzel} et~al.}{2013}]{2013MNRAS.432..336W}
{Wetzel} A.~R.,  {Tinker} J.~L.,  {Conroy} C.,   {van den Bosch} F.~C.,  2013, \mn@doi [\mnras] {10.1093/mnras/stt469}, \href {https://ui.adsabs.harvard.edu/abs/2013MNRAS.432..336W} {432, 336}

\bibitem[\protect\citeauthoryear{{Whitaker} et~al.,}{{Whitaker} et~al.}{2014}]{2014ApJ...795..104W}
{Whitaker} K.~E.,  et~al., 2014, \mn@doi [\apj] {10.1088/0004-637X/795/2/104}, \href {https://ui.adsabs.harvard.edu/abs/2014ApJ...795..104W} {795, 104}

\bibitem[\protect\citeauthoryear{{Willett} et~al.,}{{Willett} et~al.}{2013}]{2013MNRAS.435.2835W}
{Willett} K.~W.,  et~al., 2013, \mn@doi [\mnras] {10.1093/mnras/stt1458}, \href {https://ui.adsabs.harvard.edu/abs/2013MNRAS.435.2835W} {435, 2835}

\bibitem[\protect\citeauthoryear{{Woo}, {Carollo}, {Faber}, {Dekel}  \& {Tacchella}}{{Woo} et~al.}{2017}]{2017MNRAS.464.1077W}
{Woo} J.,  {Carollo} C.~M.,  {Faber} S.~M.,  {Dekel} A.,   {Tacchella} S.,  2017, \mn@doi [\mnras] {10.1093/mnras/stw2403}, \href {https://ui.adsabs.harvard.edu/abs/2017MNRAS.464.1077W} {464, 1077}

\bibitem[\protect\citeauthoryear{{Yang}, {Mo}, {van den Bosch}, {Pasquali}, {Li}  \& {Barden}}{{Yang} et~al.}{2007}]{2007ApJ...671..153Y}
{Yang} X.,  {Mo} H.~J.,  {van den Bosch} F.~C.,  {Pasquali} A.,  {Li} C.,   {Barden} M.,  2007, \mn@doi [\apj] {10.1086/522027}, \href {https://ui.adsabs.harvard.edu/abs/2007ApJ...671..153Y} {671, 153}

\bibitem[\protect\citeauthoryear{{Yang}, {Mo}  \& {van den Bosch}}{{Yang} et~al.}{2008}]{2008ApJ...676..248Y}
{Yang} X.,  {Mo} H.~J.,   {van den Bosch} F.~C.,  2008, \mn@doi [\apj] {10.1086/528954}, \href {https://ui.adsabs.harvard.edu/abs/2008ApJ...676..248Y} {676, 248}

\bibitem[\protect\citeauthoryear{{Yang}, {Mo}  \& {van den Bosch}}{{Yang} et~al.}{2009}]{2009ApJ...695..900Y}
{Yang} X.,  {Mo} H.~J.,   {van den Bosch} F.~C.,  2009, \mn@doi [\apj] {10.1088/0004-637X/695/2/900}, \href {https://ui.adsabs.harvard.edu/abs/2009ApJ...695..900Y} {695, 900}

\bibitem[\protect\citeauthoryear{{York} et~al.,}{{York} et~al.}{2000}]{2000AJ....120.1579Y}
{York} D.~G.,  et~al., 2000, \mn@doi [\aj] {10.1086/301513}, \href {https://ui.adsabs.harvard.edu/abs/2000AJ....120.1579Y} {120, 1579}

\bibitem[\protect\citeauthoryear{{Zabludoff} \& {Mulchaey}}{{Zabludoff} \& {Mulchaey}}{1998}]{1998ApJ...496...39Z}
{Zabludoff} A.~I.,  {Mulchaey} J.~S.,  1998, \mn@doi [\apj] {10.1086/305355}, \href {https://ui.adsabs.harvard.edu/abs/1998ApJ...496...39Z} {496, 39}

\bibitem[\protect\citeauthoryear{{Zu} \& {Mandelbaum}}{{Zu} \& {Mandelbaum}}{2016}]{2016MNRAS.457.4360Z}
{Zu} Y.,  {Mandelbaum} R.,  2016, \mn@doi [\mnras] {10.1093/mnras/stw221}, \href {https://ui.adsabs.harvard.edu/abs/2016MNRAS.457.4360Z} {457, 4360}

\bibitem[\protect\citeauthoryear{{de Carvalho}, {Ribeiro}, {Stalder}, {Rosa}, {Costa}  \& {Moura}}{{de Carvalho} et~al.}{2017}]{2017AJ....154...96D}
{de Carvalho} R.~R.,  {Ribeiro} A.~L.~B.,  {Stalder} D.~H.,  {Rosa} R.~R.,  {Costa} A.~P.,   {Moura} T.~C.,  2017, \mn@doi [\aj] {10.3847/1538-3881/aa7f2b}, \href {https://ui.adsabs.harvard.edu/abs/2017AJ....154...96D} {154, 96}

\bibitem[\protect\citeauthoryear{{de Carvalho}, {Costa}, {Moura}  \& {Ribeiro}}{{de Carvalho} et~al.}{2019}]{2019MNRAS.487L..86D}
{de Carvalho} R.~R.,  {Costa} A.~P.,  {Moura} T.~C.,   {Ribeiro} A.~L.~B.,  2019, \mn@doi [\mnras] {10.1093/mnrasl/slz084}, \href {https://ui.adsabs.harvard.edu/abs/2019MNRAS.487L..86D} {487, L86}

\bibitem[\protect\citeauthoryear{{de S{\'a}-Freitas} et~al.,}{{de S{\'a}-Freitas} et~al.}{2022}]{2022MNRAS.509.3889D}
{de S{\'a}-Freitas} C.,  et~al., 2022, \mn@doi [\mnras] {10.1093/mnras/stab3230}, \href {https://ui.adsabs.harvard.edu/abs/2022MNRAS.509.3889D} {509, 3889}

\bibitem[\protect\citeauthoryear{{de Vaucouleurs}}{{de Vaucouleurs}}{1963}]{1963ApJS....8...31D}
{de Vaucouleurs} G.,  1963, \mn@doi [\apjs] {10.1086/190084}, \href {https://ui.adsabs.harvard.edu/abs/1963ApJS....8...31D} {8, 31}

\bibitem[\protect\citeauthoryear{{van de Voort}, {Bah{\'e}}, {Bower}, {Correa}, {Crain}, {Schaye}  \& {Theuns}}{{van de Voort} et~al.}{2017}]{2017MNRAS.466.3460V}
{van de Voort} F.,  {Bah{\'e}} Y.~M.,  {Bower} R.~G.,  {Correa} C.~A.,  {Crain} R.~A.,  {Schaye} J.,   {Theuns} T.,  2017, \mn@doi [\mnras] {10.1093/mnras/stw3356}, \href {https://ui.adsabs.harvard.edu/abs/2017MNRAS.466.3460V} {466, 3460}

\makeatother
\end{thebibliography}









\bsp	
\label{lastpage}
\end{document}